\documentclass[conference]{IEEEtran}
\IEEEoverridecommandlockouts  %

\usepackage{iftex}
\ifPDFTeX\else\usepackage[T1]{fontenc}\fi
\ifPDFTeX\pdfsuppressptexinfo=-1\relax\fi
\usepackage{cite}
\usepackage{amsmath,amssymb,amsfonts}
\usepackage{graphicx}
\usepackage{textcomp}
\usepackage{xcolor}
\usepackage{url}
\usepackage{booktabs,array}
\usepackage{multirow}
\usepackage{enumitem}
\usepackage[most]{tcolorbox}
\usepackage{tikz}
\usetikzlibrary{fit}
\definecolor{archNew}{RGB}{253,224,196}
\definecolor{archNewEdge}{RGB}{190,85,10}
\definecolor{archBoth}{RGB}{226,226,226}
\usepackage{hyperref}
\hypersetup{colorlinks=true,linkcolor=blue,citecolor=blue,urlcolor=blue,
  pdfauthor={},pdftitle={},pdfsubject={},pdfkeywords={},pdfcreator={},pdfproducer={}}

\newtcolorbox{practitionerbox}{%
  colback=gray!10,
  colframe=gray!40,
  arc=2pt,
  boxrule=0.4pt,
  left=6pt, right=6pt, top=4pt, bottom=4pt,
  title=\textbf{Practitioner Takeaways},
  fonttitle=\normalfont\bfseries, coltitle=black,
  colbacktitle=gray!20,
  breakable
}
\graphicspath{{Figures/}}

\providecommand{\GfiveRonePending}[1]{\textbf{[MI250X R1 pending]}}

\newcommand{\appDecodeStepOverWeightBound}{5.8$\times$}

\newcommand{\appLLMFPEightAcceptedButBroken}{float8\_e4m3fn}

\newcommand{\appLLMFPEightGainDynamicDecode}{0.594$\times$}

\newcommand{\appLLMFPEightGainStaticDecode}{0.889$\times$}

\newcommand{\appLLMFPEightGainStaticPrefill}{1.12$\times$}

\newcommand{\appLLMFPEightKernelGain}{1.55$\times$}

\newcommand{\appLLMFPEightModelGain}{0.862$\times$}

\newcommand{\appLLMQuantCostlierCells}{10 of 50}

\newcommand{\appLLMQuantTaxMedian}{27.9\,$\mu$s}

\newcommand{\appSparseAmortMedian}{1.06$\times$}

\newcommand{\appSparseDenseFasterSpselFloor}{9--12 of 43}

\newcommand{\appSparseIndistinguishableSpselFloor}{10--15 of 43}

\newcommand{\appSparseModel}{Llama-3.1-8B-Instruct}

\newcommand{\appSparseOneShotDenseFaster}{43 of 43}

\newcommand{\appSparseSparseFasterSpselFloor}{19--21 of 43}

\newcommand{\appSparseSpselFloorUsed}{6.0--8.8\%}

\newcommand{\appTenantFPEightMakespanQFour}{1.19$\times$}
\newcommand{\appTenantFPEightMakespanQSixteen}{1.16$\times$}
\newcommand{\appTenantFPEightSoloQFour}{1.17$\times$}

\newcommand{\appTenantSTPHomoFour}{1.08}
\newcommand{\appTenantSTPHomoFourHwqSixteen}{1.11}

\newcommand{\attribBytesPerFlopMlp}{1.39$\times$}
\newcommand{\attribOperandRereadMlp}{2.81$\times$}

\newcommand{\captureDenomAgreementPct}{3.9\%}

\newcommand{\captureFPEightMax}{1.01}
\newcommand{\captureFPEightMin}{0.580}

\newcommand{\captureFPEightNClocked}{3}

\newcommand{\captureFPEightNOverOneFixedLayout}{1 of 7}

\newcommand{\captureSparseMax}{0.554}
\newcommand{\captureSparseMin}{0.376}
\newcommand{\captureSparseNShapes}{25}

\newcommand{\ceilFPEightHipblaslt}{1,080\,TFLOP/s}

\newcommand{\ceilFPSixteenHipblaslt}{565\,TFLOP/s}

\newcommand{\clockBracketTolerancePct}{2.0\%}

\newcommand{\clockExplainedPct}{95\%}

\newcommand{\clockRatioEight}{1.36}

\newcommand{\clockRatioSolo}{1.07}

\newcommand{\clockSustainRepeats}{2}
\newcommand{\computePartition}{SPX}

\newcommand{\cuCount}{228}
\newcommand{\cuPerXcd}{38}

\newcommand{\gemmInflationSpanMax}{44.5\%}

\newcommand{\isaArchRatio}{2.00}
\newcommand{\isaArchRatioPairs}{12}

\newcommand{\isaCodegenGainCdnaThree}{1.41--1.43$\times$}

\newcommand{\isaLatencySixteenCdnaThree}{16.9}

\newcommand{\isaLatencyThirtyTwoCdnaThree}{32.4}

\newcommand{\isaRescaledAgreement}{0.02\%}

\newcommand{\isoMoverQnSix}{0.70$\times$}

\newcommand{\isoMoverQnThree}{0.78$\times$}

\newcommand{\isoPartFPEightGainSix}{1.71$\times$}
\newcommand{\isoPartFPEightGainSolo}{1.73$\times$}

\newcommand{\isoPoverQnSix}{1.29$\times$}

\newcommand{\isoPoverQnThree}{1.31$\times$}

\newcommand{\isoQueueFPEightGainSix}{2.48$\times$}
\newcommand{\isoQueueFPEightGainSolo}{1.85$\times$}

\newcommand{\ldsPerCu}{64\,KiB}

\newcommand{\memoryPartition}{NPS1}

\newcommand{\mithreeDeviceSpreadPct}{4.0\%}

\newcommand{\operandFill}{\texttt{canonical\_rand}}

\newcommand{\selectionEffectSettled}{25\%}

\newcommand{\selectionEffectWorstFactor}{35.5$\times$}

\newcommand{\shareFPEightCaptureEight}{1.23}

\newcommand{\shareFPEightCaptureSolo}{0.958}
\newcommand{\shareFPEightCaptureTwo}{1.29}
\newcommand{\shareFPEightGainDelta}{+28.6\%}

\newcommand{\shareFPEightGainEight}{2.46$\times$}
\newcommand{\shareFPEightGainSolo}{1.92$\times$}
\newcommand{\shareFPEightRangeEight}{[2.45, 2.50]}
\newcommand{\shareFPEightRangeSolo}{[1.91, 1.92]}

\newcommand{\shareSparseGainEight}{1.39$\times$}

\newcommand{\shareTenants}{1, 2, 4, 8}
\newcommand{\smfmacArchGainRange}{1.99--2.04}

\newcommand{\smfmacClockRatioRange}{0.786--0.908}
\newcommand{\smfmacDeliveredGainRange}{1.57--1.84}
\newcommand{\smfmacPairs}{10}
\newcommand{\sparseBFSixteenAdvMeas}{17--21\%}
\newcommand{\sparseBFSixteenAdvPred}{1.5\%}

\newcommand{\sparseBFSixteenPredOverMeas}{8.7$\times$}

\newcommand{\sparseCeilQuotientFPSixteen}{0.983}

\newcommand{\sparseCompressHoldoutMaxErr}{51.2\%}

\newcommand{\sparseExploratoryCrossingSpreadFPSixteen}{4.78$\times$}

\newcommand{\sparseHoldoutCells}{7}

\newcommand{\sparseHoldoutFasterFracFPSixteen}{18 of 30}

\newcommand{\sparsePruneHoldoutMaxErr}{2.9\%}

\newcommand{\sparseSelectFlipCount}{4--8 of 18}

\newcommand{\sparseSelectFloor}{6.0--8.8\%}

\newcommand{\sparseSelectMatchedMedian}{1.21--1.23}

\newcommand{\sparseSelectNondetCells}{5 of 6}

\newcommand{\sparseShapesFasterFrac}{8 of 25}
\newcommand{\sparseShareAlgoSwapEffect}{+8.6\%}

\newcommand{\sparseShareTowardSparseCount}{4 of 6}

\newcommand{\tileFPEightMlp}{\texttt{512x144x64}}
\newcommand{\tileFPSixteenMlp}{\texttt{512x144x32}}
\newcommand{\timedRepeats}{30}
\newcommand{\warmupIters}{10}

\newcommand{\xcdCount}{6}

\newcommand{\appFusedReqGainMIThreeHundredABThirtyTwo}{1.42}

\providecommand{\GfiveRonePending}[1]{\textbf{[MI250X R1 pending]}}

\newcommand{\appLLMPromptTokens}{512}

\newcommand{\appPrefillGcdSlowdownBatchMax}{2.90$\times$}
\newcommand{\appPrefillGcdSlowdownRange}{1.97--2.94$\times$}

\newcommand{\appTenantJainTwo}{1.00}

\newcommand{\appTenantSTPTwoGcd}{1.13}

\newcommand{\appTenantSTPTwoMIThreeHundredA}{1.07}

\newcommand{\cdnaTwoCuCount}{110}

\newcommand{\cdnaTwoGcdSpreadPct}{6.71\%}

\newcommand{\cdnaTwoGcdsPerPackage}{2}

\newcommand{\cdnaTwoIsaRescaledAgreement}{0.05\%}

\newcommand{\genCapClockRatioMeasuredRange}{1.02--1.05$\times$}

\newcommand{\genCapPkgEndToEndCapBRange}{0.831--0.872}

\newcommand{\genCapPkgEndToEndGainEightK}{3.95$\times$}

\newcommand{\genCapPkgFactorOneCapBRange}{0.870--0.980}

\newcommand{\genCapPkgFactorOneGainRange}{1.20--2.15$\times$}

\newcommand{\isaDeviceRatioFPEightSixteen}{8.50$\times$}
\newcommand{\isaDeviceRatioFPEightThirtyTwo}{9.11$\times$}

\newcommand{\isaDeviceRatioFPSixteenSixteen}{4.30$\times$}
\newcommand{\isaDeviceRatioFPSixteenThirtyTwo}{4.69$\times$}

\newcommand{\isaIssueRateRatioFPSixteenSixteen}{1.78$\times$}
\newcommand{\isaIssueRateRatioFPSixteenThirtyTwo}{2.00$\times$}

\newcommand{\isaPeakClockRatio}{1.24$\times$}

\newcommand{\appFusedReqGainGcdBThirtyTwo}{1.22}

\newcommand{\sysDpmLevelMITwoHundredFiftyX}{manual}

\newcommand{\sysPowerCapMIThreeHundredA}{550 W per APU socket}
\newcommand{\sysPowerCapMITwoHundredFiftyX}{560 W per OAM}

\newcommand{\sysRocmMIThreeHundredA}{7.2.0}

\newcommand{\sysTorchMIThreeHundredA}{2.7.0+rocm6.3 (HIP 6.3.42131)}

\newcommand{\hpcAppGemmK}{2,560}

\newcommand{\hpcAppPhiGcdOneTwelveK}{1.05}

\newcommand{\hpcAppPhiMIThreeHundredAOneThirtyEightK}{0.608}

\newcommand{\hpcAppPsiGcdOneTwelveK}{1.28}

\newcommand{\hpcAppResidualThreshold}{16}
\newcommand{\hpcAppSPkg}{2.00}

\newcommand{\hpcAppSolvesPerSize}{32}

\newcommand{\hpcAppTMIThreeHundredAOneTwelveK}{3.01}

\newcommand{\hpcAppTTSRatioOneTwelveK}{2.35}

\newcommand{\hpcAppVersion}{2.1.0}
\newcommand{\hpcAppWarmupSolves}{2}

\newcommand{\refWorkDgemmGcdVsMI}{0.377}
\newcommand{\refWorkDgemmMIThreeHundredA}{69.4}

\newcommand{\refWorkTriadGcdVsMI}{0.319}
\newcommand{\refWorkTriadMIThreeHundredA}{3.26}

\newcommand{\ftzNirMI}{3}

\newcommand{\ftzNirStockRange}{4--5}

\newcommand{\ftzRTwoExactOverMIMaxDevPct}{0.43\%}

\newcommand{\ftzSubnormalFracMin}{99.9\%}

\providecommand{\MIThreeHundredAPending}[1]{\textbf{[MI300A pending]}}

\newcommand{\accCrossFPSixteenPct}{0.00570}

\newcommand{\accDecoderLinears}{224}
\newcommand{\accDeltaFPEightDynamicPctMIThreeHundredA}{0.85}
\newcommand{\accDeltaFPEightDynamicPctMIThreeHundredACI}{[0.79, 0.90]}
\newcommand{\accDeltaFPEightStaticPctMIThreeHundredA}{1.02}
\newcommand{\accDeltaFPEightStaticPctMIThreeHundredACI}{[0.96, 1.09]}

\newcommand{\accPplFPSixteen}{6.39}
\newcommand{\accPplFPSixteenMIThreeHundredA}{6.39}

\newcommand{\accPplSMagDecMIThreeHundredA}{231}

\newcommand{\accPplSTileAllMIThreeHundredA}{10,418}

\newcommand{\accRatioSMagDecMIThreeHundredA}{36.2}

\newcommand{\accRatioSTileAllMIThreeHundredA}{1630}

\newcommand{\accScoredTokens}{262,143}

\newcommand{\accPplFusedDeltaPctGcd}{$-$0.021}
\newcommand{\accPplFusedDeltaPctMIThreeHundredA}{$-$0.0001}

\newcommand{\spclkClockShapes}{5}
\newcommand{\spclkKappaMedian}{1.00}
\newcommand{\spclkKappaRange}{0.93--1.03}
\newcommand{\spclkLdsRatioMedian}{1.92$\times$}
\newcommand{\spclkRhoRange}{1.03--1.12}
\newcommand{\spclkShapes}{27 of 36}
\newcommand{\spclkUpsilonMedian}{0.46}
\newcommand{\spclkUpsilonRange}{0.39--0.60}

\providecommand{\MIThreeHundredAPending}[1]{\textbf{[MI300A pending]}}

\newcommand{\appServeFPEightCaptureDynamicRange}{0.564--0.719}
\newcommand{\appServeFPEightCaptureStaticRange}{0.580--0.689}
\newcommand{\appServeFPEightGainDynamicRange}{1.13--1.44}
\newcommand{\appServeFPEightGainStaticRange}{1.16--1.38}

\newcommand{\appServeGenGcdRange}{2.26--2.96}

\newcommand{\appServeGenPkgRange}{1.36--1.73}

\newcommand{\appServeTokMIThreeHundredAFixedSixtyFour}{2,887}

\newcommand{\appServeTtftMIThreeHundredAFixedSixtyFour}{119}
\newcommand{\appServeTtftFPEightDynMIThreeHundredAFixedSixtyFour}{79.5}

\newcommand{\appPartRatioRange}{0.78--0.96}
\newcommand{\appPartRatioGcd}{1.76}
\newcommand{\appPartSoloSlowdownCpx}{2.6}

\providecommand{\MIThreeHundredAPending}[1]{\textbf{[MI300A pending]}}

\newcommand{\appTrainFPEightGainMIThreeHundredA}{1.15}

\newcommand{\appTrainGenGcd}{3.31}

\newcommand{\appTrainGenPkg}{1.94}

\newcommand{\appTrainLossFPEightMaxRelPctMIThreeHundredA}{36\%}

\newcommand{\appTrainParamsBillion}{1.48}

\newcommand{\appTrainStepMsMIThreeHundredA}{643}

\newcommand{\appTrainTokensPerStep}{16,384}

\newcommand{\appTrainKernelRatioFPEight}{1.65}

\newcommand{\appTrainPowerCapSecondNode}{760\,W}

\newcommand{\appTrainTorchao}{0.18.0}

\newcommand{\appTrainFusedGainMIThreeHundredA}{1.08}

\makeatletter\@addtoreset{subsubsection}{section}\makeatother

\begin{document}

\title{Capability Gain from ISA to Shared Execution on AMD MI300A and MI250X}

\author{
\IEEEauthorblockN{Aaron Jarmusch}
\IEEEauthorblockA{\textit{dept. of Computer Information Sciences} \\
\textit{University of Delaware}\\
Newark, US \\
}
\and
\IEEEauthorblockN{Sunita Chandrasekaran}
\IEEEauthorblockA{\textit{dept. of Computer Information Sciences} \\
\textit{University of Delaware}\\
Newark, US \\
schandra@udel.edu}
}
\maketitle

\begin{abstract}
Each new GPU generation is showcased by its peak instruction rates, yet users hardly ever run instructions directly: they reach the hardware's performance through vendor libraries, applications and devices shared with other jobs. How much of a theoretical speedup survives these layers is rarely measured, so a team deciding whether to move to a new generation, or to adopt FP8 or 2:4 sparsity, lacks adequate guidance. AMD's MI300A (CDNA3) adds new FP8 matrix instructions and 2:4 structured-sparse instructions that the predecessor, the MI250X (CDNA2), does not have. We analyze this theoretical speedup gap on AMD's MI300A (CDNA3) against the MI250X (CDNA2) by defining a new metric, capability gain.

Capability gain is the fraction of a feature's instruction-level speedup that it keeps at a higher level, such as a library, an application or a shared device. We trace it from ISA microbenchmarks through hipBLASLt and hipSPARSELt to LLM inference and mixed-precision HPL, and onto a device shared by up to eight concurrent streams. FP8 keeps most of its speedup through the library, completing an $8192^3$ GEMM \genCapPkgEndToEndGainEightK{} faster than the MI250X with FP16, but unfused quantization turns it into a per-layer slowdown in an LLM, and 2:4 sparsity keeps at most about half of its speedup in hipSPARSELt. Under queue sharing, FP8's GEMM capability gain rises above 1. How fairly streams share the device depends on the number of hardware queues, not on the numeric precision, yet shared LLM serving gains nothing from FP8.
\end{abstract}

\begin{IEEEkeywords}
AMD MI300A, AMD MI250X, CDNA, FP8, structured sparsity, matrix cores, GPU sharing, microbenchmarking
\end{IEEEkeywords}

\section{Introduction}
\label{sec:intro}

Recent GPU generations increase matrix throughput mostly through narrower formats and structured sparsity. Deep learning uses FP8~\cite{micikevicius_fp8,ocp_fp8_spec} and 2:4 sparsity~\cite{gale_structured_sparsity}, and HPC solvers now use mixed precision~\cite{hpl_mxp,haidar_mixed_precision_ir_sc18}. Vendors advertise each feature by its peak instruction rate, but users reach it through libraries and applications. Before adopting FP8, 2:4 or a new generation, users need to know how much of that rate their own code gets.

AMD added FP8 and 2:4 structured sparsity to its CDNA3 GPUs. El Capitan, the TOP500's fastest system from November 2024 to June 2026~\cite{top500_nov2024,top500_jun2026}, is built on CDNA3's MI300A APU, and Frontier, still in production, is built on CDNA2's MI250X, which has neither feature. On paper, a CDNA3 compute unit (CU) can do $2\times$ the FP16 matrix work per cycle of a CDNA2 CU, $4\times$ with FP8, and $8\times$ with FP8 and 2:4 sparsity~\cite{amd_cdna2_whitepaper,mi200_datasheet,amd_mi300a_whitepaper}. AMD does not report how much of these rates a library, an application or a shared device keeps, or how much faster CDNA3 is than CDNA2 in practice.

The closest study, by Hanindhito and Patel~\cite{hanindhito_cdna3_family_sc25}, characterizes CDNA3 at the instruction, library and application levels on the MI300X and MI325X. They measure FP8 only on the matrix cores, measure neither sparsity nor sharing, and do not compare across generations. The one generational AMD comparison, by Tee et al.~\cite{mall_is_open_sc25}, covers only memory. As Table~\ref{tab:coverage} shows, no prior AMD study measures FP8 or 2:4 against the previous generation.

\begin{figure}[t]
  \centering
  \includegraphics[width=\columnwidth]{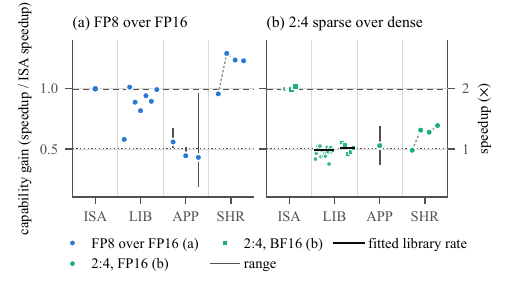}
  \caption{Capability gain of FP8 and 2:4 on the MI300A at four levels. Points: GEMM shapes in increasing size (LIB); prefill, decode and one layer (APP); 1, 2, 4 and 8 streams (SHR). Dashed: full speedup; dotted: none.}
  \label{fig:capture}
  \vspace{-1em}
\end{figure}

We track each capability with one metric, \emph{capability gain}: the speedup it delivers at a level divided by its instruction-level speedup (Section~\ref{sec:method:gain}). Fig.~\ref{fig:capture} shows it on the MI300A at four levels: the instruction set (ISA), the vendor library (LIB), the application (APP) and the shared device (SHR), where several GEMM streams run at once. Fig.~\ref{fig:capture}(a) shows that FP8 keeps \captureFPEightMin--\captureFPEightMax{} of its instruction-level speedup in hipBLASLt, but a whole LLM layer in FP8 runs at only \appLLMFPEightModelGain{} FP16's speed. Fig.~\ref{fig:capture}(b) shows that 2:4 keeps at most \captureSparseMax{} in hipSPARSELt, which is about no speedup. Sharing helps both: with eight streams, FP8's capability gain rises to \shareFPEightCaptureEight{} and sparse runs \shareSparseGainEight{} faster than dense. Each drop in Fig.~\ref{fig:capture} lines up with a specific boundary, such as the instruction clock, the library's kernel choice, an unfused conversion to FP8 or a shared queue pool, and each has a cause we can measure.

We answer four research questions. \textbf{RQ1:} Does the instruction-level speedup survive the library (Sections~\ref{sec:isa} and~\ref{sec:fp8})? \textbf{RQ2: }Where does 2:4 lose its speedup (Section~\ref{sec:sparsity})? \textbf{RQ3:} How does sharing a device change capability gain (Section~\ref{sec:cotenancy})? \textbf{RQ4: }How much of FP8's and 2:4's speedup reaches real applications (Section~\ref{sec:apps})?

To our knowledge, this is the first study to measure FP8 and 2:4 on AMD hardware against the previous generation, and the first to measure sharing on both generations under one protocol, with the number of hardware queues (4 or 16) as a controlled variable.

In summary, this paper makes the following contributions:

\begin{itemize}[leftmargin=*,itemsep=1pt,topsep=2pt]
\item We define \emph{capability gain} and a method to trace it across four levels (instruction, library, application and shared device) on two generations, under a protocol that controls kernel selection and operand values, and we attribute each loss to a mechanism with hardware counters and measured clocks (Section~\ref{sec:method}, Fig.~\ref{fig:capture}).
\item We compare CDNA3's FP8 matrix path on the MI300A with CDNA2's FP16 path on the MI250X, from instruction microbenchmarks through hipBLASLt, per package and per CU$\cdot$clock (Sections~\ref{sec:isa} and~\ref{sec:fp8}).
\item We evaluate CDNA3's 2:4 sparsity stage by stage, from the SMFMAC instructions through hipSPARSELt, and test a break-even model on shapes not used to fit it (Section~\ref{sec:sparsity}).
\item We present a multi-stream benchmark for GPUs and run it on both generations under one protocol, varying streams, hardware queues, CU masks and partitions (Section~\ref{sec:cotenancy}).
\item We measure FP8 and 2:4 in LLM inference, serving and training and in mixed-precision HPL, and derive guidance for library developers, system administrators and users (Sections~\ref{sec:apps} and~\ref{sec:discussion}).
\end{itemize}

We will open source the benchmarks upon acceptance; we withhold them now for double-blind review.

\section{Background: CDNA3 and CDNA2}
\label{sec:background}

CDNA3 adds FP8 matrix instructions, 2:4 structured sparsity and compute partitions to CDNA2's design, and like CDNA2, it dispatches concurrent work through hardware queues.

\begin{table}[t]
\centering\scriptsize
\caption{Architectural parameters of the MI300A and MI250X.}
\label{tab:gen-spec}
\setlength{\tabcolsep}{3pt}
\begin{tabular}{@{}lll@{}}
\toprule
 & MI300A (CDNA3) & MI250X (CDNA2) \\
\midrule
Dies & 6 XCDs & 2 GCDs \\
CUs & 228 (38 per XCD) & 220 (110 per GCD) \\
Peak engine clock & 2.1\,GHz & 1.7\,GHz \\
LDS per CU & 64\,KiB & 64\,KiB \\
FP16 FLOP/CU/clock & 2,048 & 1,024 \\
FP8 FLOP/CU/clock & 4,096 & -- \\
Peak FP16 matrix & 981\,TFLOP/s & 383\,TFLOP/s \\
Peak FP8 matrix & 1,961\,TFLOP/s & -- \\
Peak 2:4 FP16 & 1,961\,TFLOP/s & -- \\
Peak 2:4 FP8 & 3,922\,TFLOP/s & -- \\
\bottomrule
\end{tabular}
\end{table}

\begin{figure}[t]
    \centering
    {\scriptsize (a)}\hspace{4pt}%
    \includegraphics[width=0.55\columnwidth]{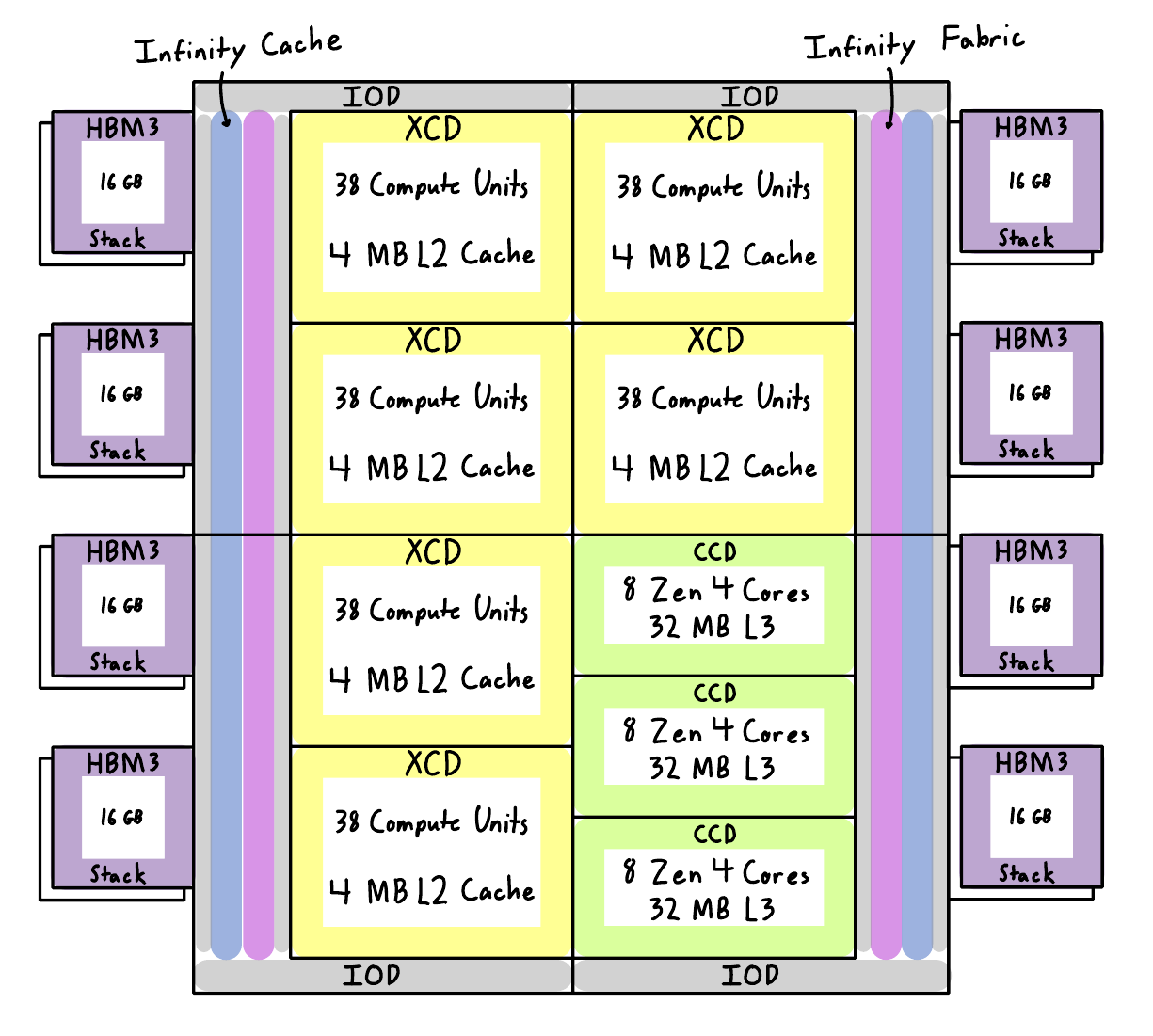}\\[3pt]
    {\scriptsize (b)}\hspace{4pt}%
    \resizebox{0.92\columnwidth}{!}{\begin{tikzpicture}[
    font=\scriptsize,
    >=stealth,
    blk/.style={draw, rounded corners=1pt, align=center, inner sep=2pt,
                minimum height=0.46cm},
    new/.style={blk, fill=archNew, draw=archNewEdge, line width=0.6pt},
    both/.style={blk, fill=archBoth},
    plain/.style={blk, fill=white},
    sec/.style={font=\scriptsize\bfseries, text=archNewEdge},
  ]
  \node[plain, minimum width=2.2cm] (streams) at (1.2,3.05) {HIP streams};
  \node[both,  minimum width=2.2cm] (hwq)     at (1.2,1.95) {Hardware queues\\(4 or 16)};
  \node[plain, minimum width=2.2cm] (ace)     at (1.2,0.90) {ACEs};
  \draw[->] (streams) -- (hwq);
  \draw[->] (hwq) -- (ace);
  \node[draw, dashed, rounded corners=2pt, inner sep=3pt, fit=(streams)(hwq)(ace)] (part) {};
  \node[sec, anchor=west, inner sep=1pt] at (part.east |- hwq) {\S\ref{sec:cotenancy}};
  \node[anchor=north, align=center, font=\tiny] at (part.south)
       {Partition: SPX / TPX / CPX\\(\computePartition{} unless stated)};

  \node[draw, thick, rounded corners=2pt, minimum width=4.85cm, minimum height=3.55cm,
        anchor=south west] (cu) at (3.3,-0.05) {};
  \node[anchor=north west, font=\scriptsize\bfseries] at (cu.north west)
       {Compute Unit (1 of \cuCount)};
  \draw[->] (ace.east) -- (cu.west |- ace.east);

  \node[anchor=west, font=\tiny] at (3.45,2.95) {Matrix core};
  \node[both, minimum width=3.4cm, anchor=west] (f16) at (3.45,2.55) {FP16/BF16 MFMA};
  \node[new,  minimum width=3.4cm, anchor=west] (f8)  at (3.45,2.03) {FP8/BF8 MFMA};
  \node[new,  minimum width=3.4cm, anchor=west] (sp)  at (3.45,1.51) {2:4 sparse SMFMAC};
  \node[sec, anchor=west] at (f16.east) {\S\ref{sec:isa}};
  \node[sec, anchor=west] at (f8.east)  {\S\ref{sec:fp8}};
  \node[sec, anchor=west] at (sp.east)  {\S\ref{sec:sparsity}};

  \foreach \i in {0,1,2,3} {
    \node[plain, minimum width=0.78cm, anchor=west] at (3.45+0.86*\i,0.86) {SIMD};
  }
  \node[plain, minimum width=3.4cm, anchor=west] (lds) at (3.45,0.30)
       {LDS (\ldsPerCu)};

  \node[new, minimum height=0.2cm, minimum width=0.3cm, inner sep=0pt] (k1) at (0.35,-0.60) {};
  \node[anchor=west, font=\tiny] at (k1.east) {new in CDNA3};
  \node[both, minimum height=0.2cm, minimum width=0.3cm, inner sep=0pt] (k2) at (2.35,-0.60) {};
  \node[anchor=west, font=\tiny] at (k2.east) {on CDNA2 and CDNA3};
  \node[sec, anchor=west, font=\tiny\bfseries] at (4.9,-0.60) {\S\,= section it's mentioned};
\end{tikzpicture}}
    \caption{MI300A architecture. (a) The full package. (b) One compute unit,
    and the dispatch path from HIP streams through hardware queues to it.}
    \label{fig:mi300a-arch}
\end{figure}

The MI300A (Fig.~\ref{fig:mi300a-arch}) has \xcdCount{} CDNA3 accelerator dies (XCDs) of \cuPerXcd{} enabled CUs each, \cuCount{} in total, and its CPU cores share the GPU's HBM3~\cite{amd_mi300a_whitepaper}.%
 The MI250X (CDNA2) is an Open Compute Project (OCP) Accelerator Module (OAM) package of \cdnaTwoGcdsPerPackage{} graphics compute dies (GCDs) with \cdnaTwoCuCount{} CUs each~\cite{amd_cdna2_whitepaper}. Table~\ref{tab:gen-spec} lists the vendor peaks~\cite{mi300a_datasheet,mi200_datasheet}.

\textbf{FP8 matrix instructions.} Both generations issue matrix work as matrix fused multiply-add (MFMA) instructions~\cite{cdna3_isa_guide}. CDNA3 adds FP8 and BF8 MFMA instructions that accumulate in FP32, and the datasheet states that they run at twice the FP16 rate~\cite{mi300a_datasheet}. Their encodings are the FNUZ variants of E4M3 and E5M2%
, which have no infinities and no negative zero~\cite{cdna3_isa_guide}. These differ from the OCP 8-bit floating-point (OFP8) formats~\cite{micikevicius_fp8,ocp_fp8_spec}, so software must support FNUZ explicitly; as Table~\ref{tab:gen-support} shows, rocBLAS has no FP8 type.

\textbf{2:4 structured sparsity.} The sparse MFMA instructions (SMFMAC) execute 2:4 sparsity, in which two of every four consecutive elements are zero~\cite{cdna3_isa_guide,amd_mi300a_whitepaper}, and the datasheet states that they run at twice the dense rate~\cite{mi300a_datasheet}. The saving is in arithmetic only. Pruning and compression are separate stages the application must run before the multiply, and Table~\ref{tab:gen-support} shows that hipSPARSELt does not expose every type that SMFMAC accepts. The datasheet states neither the cost of those stages nor the reuse at which the $2\times$ pays for them, so we time each stage in Section~\ref{sec:sparsity}.

\textbf{Partitions and hardware queues.} Users can split the MI300A into SPX, TPX or CPX compute partitions (one, three or six partitions)~\cite{amd_mi300a_whitepaper,amr_partitioning}. The MI250X has no partition modes; its GCDs enumerate as separate devices, so the GCD is the finest unit a scheduler can allocate. On both parts, hardware queues map onto asynchronous compute engines~\cite{amd_mi300a_whitepaper,hsa_runtime_spec}.  Section~\ref{sec:cotenancy} measures them with precision and sparsity varied.

\providecommand{\cyes}{$\checkmark$}
\providecommand{\cno}{--}

\section{Related Work}
\label{sec:related}

\begin{table}[t]
  \centering
  \caption{Coverage of prior microbenchmark
  studies and this paper. Sh.: concurrent streams; Gen.: two generations.}
  \label{tab:coverage}
  \scriptsize
  \setlength{\tabcolsep}{2.2pt}
  \begin{tabular}{@{}llccccccc@{}}
    \toprule
    Work & Platform & ISA & LIB & APP & FP8 & 2:4 & Sh. & Gen. \\
    \midrule
    Schieffer~\cite{schieffer_amd_matrix_cores_ispass24}            & MI250X        & \cyes  & \cyes  & \cno   & \cno   & \cno   & \cno   & \cno  \\
    Hanindhito~\cite{hanindhito_cdna3_family_sc25}                  & MI300X/325X   & \cyes  & \cyes  & \cyes  & \cyes  & \cno   & \cno   & \cno  \\
    Jarmusch~\cite{blackwell_microbench}                            & B200, H200    & \cyes  & \cyes  & \cyes  & \cyes  & \cyes  & \cno   & \cyes \\
    Tee~\cite{mall_is_open_sc25}                                  & MI250X/300A/X & \cyes  & \cno   & \cno   & \cno   & \cno   & \cno   & \cyes \\
    \midrule
    \textbf{Our paper}                                             & \textbf{MI300A, MI250X} & \cyes & \cyes & \cyes & \cyes & \cyes & \cyes & \cyes \\
    \bottomrule
  \end{tabular}
\end{table}

No prior AMD work in Table~\ref{tab:coverage} covers more than four of its seven dimensions, and none pairs FP8 or 2:4 with a generational comparison.

\subsubsection{Instruction and library level}
Schieffer et al.~\cite{schieffer_amd_matrix_cores_ispass24} characterize the MI250X's CDNA2 matrix cores with microbenchmarks and rocBLAS GEMM in FP64, FP32 and mixed FP16. We re-measure the dense path so that both sides of a generational ratio come from one protocol. Hanindhito and Patel~\cite{hanindhito_cdna3_family_sc25} characterize the MI300X and MI325X at the instruction, library and application levels. They leave 8-bit GEMM out because of a hipBLASLt performance regression, and they measure no sparsity, sharing or CDNA2 part. Ambati and Diep~\cite{ambati_mi300x_analysis_arxiv} correct FP8 and FP16 library GEMM efficiency on the MI300X for the measured clock, as capability gain~$B$ does. They do so at one level and on one generation. NVIDIA parts have comparable studies~\cite{blackwell_microbench,hopper_benchmark,demystify_gpu_micro,gpu_microbenchmarking}.

\subsubsection{Memory and partitioning}
Tee et al.~\cite{mall_is_open_sc25} compare the memory-latency hierarchy of MI250X, MI300A and MI300X, including the last-level cache, memory (NPS) partitioning and contention. Their generational comparison is for memory and ours is for the compute features CDNA3 adds, so we do not re-measure memory. Prior work~\cite{wahlgren_mi300a_upm,kuncham_one_memory_many_paths_ipdps26,antepara_roofline_superchips_sc25,schieffer_mi300a_interapu} measures MI300A unified memory, copy strategies, CPU--GPU contention and inter-APU communication, and Abouelmagd et al.~\cite{amr_partitioning} sweep the partition modes.

\subsubsection{Sharing}
Otterness and Anderson study CU masking~\cite{otterness_amd_ecrts20} and the path from HIP streams through user-mode queues to the asynchronous compute engines~\cite{otterness_amd_scheduling} on GCN GPUs, and show that the runtime multiplexes all streams onto a small pool of hardware queues. ECLIP~\cite{quach_eclip_islped25} uses CU-masked stream pools on MI50, and RAPID-Serve~\cite{rapid_serve} co-locates LLM prefill and decode with CU masking on MI300X. None of the three varies precision or sparsity or compares two generations. We do both, with the queue count as a controlled variable.

\subsubsection{Sparsity}
Many studies cover 2:4 sparsity on NVIDIA GPUs~\cite{structured_sparsity,gale_structured_sparsity,2_4_sparse_training,continuous_pruning_2_4,pytorch_2_4_sparsity}. On AMD, Dickmann et al.~\cite{dickmann_cug25} compare a 2:4-sparse LLM on MI300A with a larger dense model, which confounds sparsity with model size, and do not report whether the sparse hardware path runs. To our knowledge, no AMD study separates pruning and compression from the multiply or reports a break-even point. We time each hipSPARSELt stage, test a break-even model on held-out shapes, and run the hardware 2:4 path under sharing.

\section{Methodology}
\label{sec:method}
\subsection{Hardware setup}
\label{sec:method:hw}

We run on two systems with nodes for MI300A and MI250X to compare package against package. The MI300A runs in \computePartition{}/\memoryPartition{} mode with a power cap of \sysPowerCapMIThreeHundredA, and the MI250X is capped at \sysPowerCapMITwoHundredFiftyX. A one-GCD run leaves the sibling GCD idle, and a package run uses both. We compare packages because the GCDs share one power cap and only one reports the package's power; one GCD is a secondary comparison.

\subsection{Experiment methodology}
\label{sec:method:recipes}

We use one benchmark per level, from single matrix instructions to whole applications, and Sections~\ref{sec:isa}--\ref{sec:apps} report their results in this order. All run on ROCm \sysRocmMIThreeHundredA, the LLM benchmarks use PyTorch \sysTorchMIThreeHundredA, and the artifact lists the library, driver and OS versions.

\subsubsection{Matrix-instruction microbenchmarks}
Our microbenchmarks time single matrix instructions with no library involved. Each issues one MFMA or SMFMAC compiler builtin in a loop whose operands stay in registers. Throughput variants run independent accumulator chains in every wavefront across the device; latency variants run one wavefront through a chain of dependent MFMAs. Each wavefront reads its shader clock and a constant-rate counter around the loop, which yields cycles per instruction. We build each microbenchmark with the default code generation and with a form that keeps accumulators in the matrix registers and inspect the disassembly.

\subsubsection{Library benchmarks}
We time each GEMM in hipBLASLt, rocBLAS and hipSPARSELt with its fastest kernel, so the results reflect the library and not its heuristic. The shapes are $N^3$ for $N$ from 1024 to 16384 in powers of two, an MLP shape ($M{\times}N{\times}K = 4096{\times}16384{\times}17408$) and a thin-$K$ shape ($8192{\times}8192{\times}1024$). Every GEMM uses the NN layout and FP32 accumulation, FP8 GEMMs write FP16, and throughput is $2MNK$ over the median time. We compare two libraries only when the input, output and accumulator datatypes all match. For hipBLASLt we time every algorithm that \texttt{getAllAlgos} returns for those datatypes and keep the fastest; rocBLAS runs \texttt{rocblas\_gemm\_ex} with its standard algorithm. We time hipSPARSELt's prune, compress, plan and matmul stages separately, and run the sparse matmul and the matching dense GEMM at their best algorithms.

\subsubsection{Multi-stream benchmark}
The multi-stream benchmark runs several GEMM streams on one device at once and varies how many hardware queues they share. Each HIP stream runs the same GEMM once per round, always with the same kernel, chosen before the run. We vary the stream count over \shareTenants, start the streams together with a flag set on the GPU, and shuffle their launch order with a recorded seed. HIP maps streams onto a default pool of four hardware queues~\cite{otterness_amd_scheduling}, so from four streams up, streams share queues. We run every configuration with this default and with 16 queues, where no two streams share one. To divide the MI300A among streams that do the same total work, we also limit each stream to its own CUs with CU masks, or split the device into three or six separate devices with TPX and CPX partitions.

\subsubsection{Application benchmarks}
The application benchmarks measure FP8 and 2:4 in one LLM, in inference, serving and training, and the FP16 matrix path in mixed-precision HPL.

\textbf{LLM.} We run \appSparseModel{} with \appLLMPromptTokens-token prompts on a plain PyTorch/transformers harness that keeps every kernel visible, and time each phase (prefill and decode) and each layer. Attention uses PyTorch's MATH backend, checked against an FP32 reference before each run; fused-attention (FLASH) runs are reported beside it. We time 2:4 layers in hipSPARSELt against a tuned dense baseline at every linear shape, on the MI300A only, and measure WikiText-2 perplexity over \accScoredTokens{} tokens with the same masks on both machines, without fine-tuning. We serve the model in three ways: with vLLM~\cite{vllm_sosp23}, which fuses activation quantization (FP16 and FP8 on the MI300A, FP16 on the MI250X GCD); as several processes with their own replicas on one device; and as \xcdCount{} processes in CPX partitions against the same processes in SPX. Serving reports throughput, time to first token and time per output token over a fixed request trace and a mixed-length one. For training, we measure the time per step of a GPT-style model with \appTrainParamsBillion{} billion parameters at \appTrainTokensPerStep{} tokens per step. The MI300A trains it in BF16 and in FP8, with torchao \appTrainTorchao{} converting the linear layers to FP8 with dynamic per-tensor scaling; the MI250X GCD and package train it in BF16. Every run starts from the same weights and data order, and we compare the losses step by step. The training runs use an MI300A node capped at \appTrainPowerCapSecondNode.

\textbf{HPC.} We run mixed-precision HPL~\cite{hpl_mxp,haidar_mixed_precision_ir_sc18} in AMD's unmodified rocHPL-MxP \hpcAppVersion~\cite{rochpl_mxp}. Each solve of a dense system $Ax=b$ factors $A$ with a rocBLAS FP16 GEMM in its LU trailing update ($K=\hpcAppGemmK$) and refines the result in FP64; it passes only if its scaled FP64 residual falls below \hpcAppResidualThreshold. Each device runs one MPI rank, and the package one solve per GCD. We time \hpcAppSolvesPerSize{} solves per size with the benchmark's host clock and discard the first \hpcAppWarmupSolves. We also run a STREAM Triad and rocBLAS DGEMM.

\subsection{Capability gain}
\label{sec:method:gain}

Capability gain measures how much of a capability's instruction-level speedup a level keeps. For a capability (FP8 over FP16, or 2:4 sparse over dense) at level $\ell$, it is
\begin{equation}
  G_\ell = S_\ell \,/\, S_{\mathrm{ISA}},
  \label{eq:gain}
\end{equation}
where $S_\ell$ is the measured speedup at level $\ell$ and $S_{\mathrm{ISA}}$ is the instruction-level speedup. $G_\ell = 1$ means the level keeps the whole speedup, and $G_\ell > 1$ means it is faster than the instructions predict. We take $S_{\mathrm{ISA}}$ as the architectural ratio, \isaArchRatio, because the measured FP8/FP16 clock ratio varies with the operands (Section~\ref{sec:fp8}). Across generations, capability gain~$B$ divides a generational speedup by the architectural FLOP-per-CU-per-cycle ratio, the CU-count ratio and the measured clock ratio.

\subsection{Measurements}
\label{sec:method:stats}
\label{sec:method:validity}

Every result is the median of repeated runs without profiling. HIP events time each measured region, which ends with a stream synchronization, after at least \warmupIters{} warmup iterations and over at least \timedRepeats{} stored repeats, and we report the interquartile range with each median. We randomize the configuration order with a recorded seed. We use hardware counters only for diagnosis and for capability gain~$B$, because profiled runs serialize dispatches and inflate clocks for short kernels. We take $B$'s clock from each XCD's hardware counter of busy cycles, check it against a second such counter in the same pass, and report $B$ only at shapes where the two agree within \clockBracketTolerancePct{} on both GPUs. We normalize throughput only by measured peaks, the highest sustained throughput in a dedicated sweep, which Table~\ref{tab:gen-ceilings} lists. Library throughput varies by up to \mithreeDeviceSpreadPct{} between two MI300A devices and by up to \cdnaTwoGcdSpreadPct{} between the two GCDs of one MI250X, so we treat any smaller difference as noise.

The values in the input matrices change GEMM throughput and the MI250X's delivered clock (Section~\ref{sec:fp8}), so we fix them: every measured peak multiplies the same random input values (\operandFill), generated from a fixed random-number seed. We set the MI250X's power-management (DPM) level to \sysDpmLevelMITwoHundredFiftyX{} and measure its delivered clock inside the kernel with extra wavefronts that read the clock, not from the driver. The MI300A runs under default power management.

\begin{table}[t]
\centering
\caption{How FP8 and 2:4 are reached at each level on the MI300A, and where the MI250X
path stops.}
\label{tab:gen-support}\label{tab:support}
\scriptsize
\setlength{\tabcolsep}{3pt}
\renewcommand{\arraystretch}{1.1}
\newcommand{\ub}{\_\allowbreak}
\begin{tabular}{@{}l>{\raggedright\arraybackslash}p{0.39\columnwidth}>{\raggedright\arraybackslash}p{0.39\columnwidth}@{}}
\toprule
 & FP8 GEMM & 2:4 sparse GEMM \\
\midrule
\multicolumn{3}{@{}l}{\textit{MI300A (CDNA3)}}\\
ISA & \texttt{v\ub mfma\ub f32\ub 32x32x16\ub fp8\ub fp8} & \texttt{v\ub smfmac\ub f32\ub 32x32x16\ub f16} \\
Library & hipBLASLt with \texttt{HIP\ub R\ub 8F\ub E4M3\ub FNUZ}; rocBLAS has no FP8 type & hipSPARSELt prune $\rightarrow$ compress $\rightarrow$ plan $\rightarrow$ matmul, for FP16, BF16 and INT8; FP8 error \\
PyTorch & \texttt{torch.\ub scaled\ub mm}, \texttt{float8\ub e4m3fnuz}; OCP \texttt{e4m3fn} returns NaN & -- \\
\midrule
\multicolumn{3}{@{}l}{\textit{MI250X (CDNA2): where the path stops}}\\
ISA & compile error (no \texttt{fp8-insts}) & backend cannot select SMFMAC \\
Library & hipBLASLt: no FP8 algorithm & hipSPARSELt: errors for every type \\
PyTorch & \texttt{\ub scaled\ub mm} is MI300+ only & -- \\
\bottomrule
\end{tabular}
\end{table}

\begin{table*}[t]
  \centering
  \caption{Measured GEMM peaks and HBM bandwidth of the MI300A, one MI250X GCD and the MI250X package. Best measured rate per library; HBM: device-to-device copy kernel; \emph{per CU$\cdot$clk}: the GCD's ratio per CU and peak clock.}
  \label{tab:gen-ceilings}
  \footnotesize
  \setlength{\tabcolsep}{4pt}
  \begin{tabular}{ll rr rrrr rrr}
    \toprule
    & & \multicolumn{2}{c}{MI300A} & \multicolumn{4}{c}{MI250X, one GCD} & \multicolumn{3}{c}{MI250X package} \\
    \cmidrule(lr){3-4}\cmidrule(lr){5-8}\cmidrule(lr){9-11}
    Prec. & Library & TFLOP/s & \% peak & TFLOP/s & \% peak & vs MI300A & per CU$\cdot$clk & TFLOP/s & \% peak & vs MI300A \\
    \midrule
    \multicolumn{2}{l}{Device} & \multicolumn{2}{c}{228 CUs, 2.1\,GHz} & \multicolumn{4}{c}{110 CUs, 1.7\,GHz} & \multicolumn{3}{c}{220 CUs, 1.7\,GHz} \\
    \midrule
    FP64 & rocBLAS & 69.4 & 57\% & 26.1 & 55\% & 0.377$\times$ & 0.965$\times$ & 50.7 & 53\% & 0.731$\times$ \\
    FP64 & hipBLASLt & 16.3 & 13\% & 6.3 & 13\% & 0.386$\times$ & 0.989$\times$ & 12.7 & 13\% & 0.778$\times$ \\
    FP32 & rocBLAS & 91.7 & 75\% & 40.4 & 84\% & 0.441$\times$ & 1.13$\times$ & 78.0 & 81\% & 0.850$\times$ \\
    FP32 & hipBLASLt & 94.1 & 77\% & 40.1 & 84\% & 0.427$\times$ & 1.09$\times$ & 75.3 & 79\% & 0.800$\times$ \\
    FP16 & rocBLAS & 566 & 58\% & 128 & 67\% & 0.226$\times$ & 0.578$\times$ & 244 & 64\% & 0.432$\times$ \\
    FP16 & hipBLASLt & 565 & 58\% & 142 & 74\% & 0.251$\times$ & 0.641$\times$ & 273 & 71\% & 0.483$\times$ \\
    BF16 & rocBLAS & 590 & 60\% & 139 & 73\% & 0.236$\times$ & 0.603$\times$ & 262 & 68\% & 0.444$\times$ \\
    BF16 & hipBLASLt & 590 & 60\% & 153 & 80\% & 0.260$\times$ & 0.666$\times$ & 287 & 75\% & 0.486$\times$ \\
    FP8 & hipBLASLt & 1,080 & 55\% & \multicolumn{4}{c}{--} & \multicolumn{3}{c}{--} \\
    \midrule
    HBM & TB/s & 3.24 & 61\% & 1.17 & 71\% & 0.360$\times$ & -- & 2.34 & 71\% & 0.721$\times$ \\
    \bottomrule
  \end{tabular}
\end{table*}

\section{Instruction Level: Matrix Cores}
\label{sec:isa}

\begin{table*}[t]
  \centering
  \caption{Matrix-instruction throughput, issue cycles and dependent-issue latency on CDNA3 and CDNA2; CDNA3: the whole MI300A; CDNA2: one GCD.
  FLOP/CU/cyc at the peak clock; issue cycles at one wave per SIMD; latency over a
  32-deep dependent chain.}
  \label{tab:gen-isa}
  \footnotesize
  \setlength{\tabcolsep}{4pt}
  \begin{tabular}{lllrrrrrrr}
    \toprule
    Arch. & Prec. & Tile & TFLOP/s & \% ISA peak & \% vendor & FLOP/CU/cyc & Issue cyc & Latency cyc & vs CDNA2 FP16 \\
    \midrule
    CDNA3 & FP8 & 16x16x32 & 1,520 & 95.3\% & 77.5\% & 3,175 & 18.0 & 16.9 & 8.50$\times$ \\
    CDNA3 & FP8 & 32x32x16 & 1,596 & 100\% & 81.4\% & 3,334 & 32.0 & 32.4 & 9.11$\times$ \\
    \midrule
    CDNA3 & FP16 & 16x16x16 & 769 & 93.6\% & 78.5\% & 1,607 & 18.0 & 16.9 & 4.30$\times$ \\
    CDNA3 & FP16 & 32x32x8 & 822 & 100\% & 83.9\% & 1,718 & 32.0 & 32.4 & 4.69$\times$ \\
    \midrule
    CDNA3 & BF16 & 16x16x16 & 778 & 95.3\% & 79.4\% & 1,626 & 18.0 & 16.9 & 4.35$\times$ \\
    CDNA3 & BF16 & 32x32x8 & 816 & 100\% & 83.3\% & 1,705 & 32.0 & 32.4 & 4.66$\times$ \\
    \midrule
    CDNA2 & FP16 & 16x16x16 & 179 & 100\% & 93.4\% & 956 & 32.0 & 32.5 & 1.00$\times$ \\
    CDNA2 & FP16 & 32x32x8 & 175 & 98.0\% & 91.5\% & 937 & 64.0 & 64.1 & 1.00$\times$ \\
    \midrule
    CDNA2 & BF16 & 16x16x16 & 179 & 100\% & 93.4\% & 956 & 32.0 & 32.5 & 1.00$\times$ \\
    CDNA2 & BF16 & 32x32x8 & 175 & 97.9\% & 91.4\% & 936 & 64.0 & 64.1 & 0.999$\times$ \\
    \bottomrule
  \end{tabular}
\end{table*}

RQ1 starts at the instructions, and there the speedup survives: CDNA3 delivers AMD's advertised speedups per issue cycle, and later sections measure against them. We rescale the rates by each GPU's measured shader clock. The rescaled rates match the datasheet peaks~\cite{mi300a_datasheet,amd_cdna2_whitepaper,mi200_datasheet} to within \isaRescaledAgreement{} on CDNA3 and \cdnaTwoIsaRescaledAgreement{} on CDNA2, so any shortfall from the vendor peak lies in the clock. Table~\ref{tab:gen-isa} shows that one MI300A runs FP16 MFMA \isaDeviceRatioFPSixteenSixteen--\isaDeviceRatioFPSixteenThirtyTwo{} and FP8 MFMA \isaDeviceRatioFPEightSixteen--\isaDeviceRatioFPEightThirtyTwo{} faster than one MI250X GCD runs FP16.

\subsubsection{Tile size}
AMD's claim that CDNA3 doubles CDNA2's FP16 rate per CU per cycle~\cite{amd_mi300a_whitepaper} holds only for 32$\times$32 tiles. CDNA3 issues at \isaIssueRateRatioFPSixteenThirtyTwo{} CDNA2's rate for 32$\times$32 tiles but at only \isaIssueRateRatioFPSixteenSixteen{} for 16$\times$16 tiles, so kernels that use 16$\times$16 tiles get less than the advertised doubling. FP8 has no such limit: one FP8 MFMA instruction performs \isaArchRatio$\times$ the FLOPs of an FP16 MFMA instruction in the same number of issue cycles for every tile size (\isaArchRatioPairs{} matched pairs), so FP8's instruction-level speedup does not depend on which tile the library picks.

\subsubsection{Code generation}
The compiler sets CDNA3's measured 16$\times$16 instruction peak, so we report the best build. The default build wraps CDNA3's MFMAs in extra instructions that move accumulators, and the register-form build, which keeps accumulators in the matrix registers, runs \isaCodegenGainCdnaThree{} faster at 16$\times$16.

\subsubsection{Latency and occupancy}
FP8 and FP16 MFMA instructions of the same tile have the same latency, so FP8's $2\times$ comes from doing twice the work per instruction, not from finishing sooner. On both generations, a dependent MFMA waits about one issue interval (the cycles one instruction occupies the matrix pipe): \isaLatencySixteenCdnaThree{} and \isaLatencyThirtyTwoCdnaThree{} cycles for 16$\times$16 and 32$\times$32 tiles on CDNA3. A chain of dependent MFMAs therefore stalls little. CDNA3 reaches its measured FP8 and FP16 peaks with one wavefront per SIMD.

\section{FP8 Through the Library and Across a Generation}
\label{sec:fp8}

The answer to RQ1 is mostly yes: hipBLASLt keeps most of FP8's instruction-level \isaArchRatio$\times$ except in small GEMMs, and the library's kernel choice, not the hardware, causes the loss. An $8192^3$ GEMM runs \genCapPkgEndToEndGainEightK{} faster in FP8 on the MI300A than in FP16 on the MI250X package, close to what the hardware predicts. Table~\ref{tab:gen-library} lists each library path; CDNA2's baseline is its FP16 path at the same shape, datatypes and input values, with BF16 as an alternative.

\begin{table}[t]
  \centering
  \caption{Library GEMM throughput and share of each part's measured peak (\% meas.), MI300A vs one GCD. Lt: hipBLASLt, rB: rocBLAS.}
  \label{tab:gen-library}
  \footnotesize
  \setlength{\tabcolsep}{3.5pt}
  \renewcommand{\arraystretch}{0.92}
  \begin{tabular}{llrrrrr}
    \toprule
    & & \multicolumn{2}{c}{MI300A} & \multicolumn{2}{c}{MI250X GCD} & vs MI250X \\
    \cmidrule(lr){3-4}\cmidrule(lr){5-6}
    Shape & Path & TFLOP/s & \% meas. & TFLOP/s & \% meas. & Lt FP16 \\
    \midrule
    $1024^3$ & Lt~BF16 & 149 & 25.3 & 54.1 & 35.3 & 2.61$\times$ \\
     & Lt~FP16 & 138 & 24.3 & 57.1 & 40.3 & 2.41$\times$ \\
     & Lt~FP8 & 160 & 15.2 & \multicolumn{2}{c}{--} & 2.79$\times$ \\
     & rB~FP16 & 85.1 & 15.0 & 59.7 & 46.7 & 1.49$\times$ \\
    \addlinespace[1pt]
    $2048^3$ & Lt~BF16 & 285 & 48.4 & 102 & 66.4 & 2.91$\times$ \\
     & Lt~FP16 & 261 & 46.2 & 98.0 & 69.2 & 2.67$\times$ \\
     & Lt~FP8 & 530 & 50.5 & \multicolumn{2}{c}{--} & 5.41$\times$ \\
     & rB~FP16 & 212 & 37.4 & 99.5 & 77.8 & 2.16$\times$ \\
    \addlinespace[1pt]
    $4096^3$ & Lt~BF16 & 529 & 89.6 & 131 & 85.2 & 4.31$\times$ \\
     & Lt~FP16 & 527 & 93.2 & 123 & 86.6 & 4.29$\times$ \\
     & Lt~FP8 & 936 & 89.2 & \multicolumn{2}{c}{--} & 7.63$\times$ \\
     & rB~BF16 & 326 & 55.3 & 125 & 89.6 & 2.66$\times$ \\
     & rB~FP16 & 304 & 53.7 & 119 & 93.1 & 2.48$\times$ \\
    \addlinespace[1pt]
    thin-$K$ & Lt~BF16 & 476 & 80.7 & 117 & 76.0 & 3.95$\times$ \\
     & Lt~FP16 & 460 & 81.4 & 121 & 85.2 & 3.82$\times$ \\
     & Lt~FP8 & 753 & 71.8 & \multicolumn{2}{c}{--} & 6.25$\times$ \\
     & rB~FP16 & 415 & 73.3 & 103 & 80.4 & 3.44$\times$ \\
    \addlinespace[1pt]
    $8192^3$ & Lt~BF16 & 581 & 98.5 & 146 & 95.4 & 4.37$\times$ \\
     & Lt~FP16 & 557 & 98.6 & 133 & 94.0 & 4.19$\times$ \\
     & Lt~FP8 & 1,050 & 100 & \multicolumn{2}{c}{--} & 7.89$\times$ \\
     & rB~FP16 & 456 & 80.6 & 126 & 98.2 & 3.43$\times$ \\
    \addlinespace[1pt]
    MLP & Lt~BF16 & 591 & 100 & 150 & 97.8 & 4.37$\times$ \\
     & Lt~FP16 & 576 & 102 & 135 & 95.7 & 4.25$\times$ \\
     & Lt~FP8 & 1,032 & 98.4 & \multicolumn{2}{c}{--} & 7.62$\times$ \\
     & rB~FP16 & 496 & 87.7 & 119 & 93.0 & 3.66$\times$ \\
    \addlinespace[1pt]
    $16384^3$ & Lt~BF16 & 601 & 102 & 152 & 99.3 & 4.33$\times$ \\
     & Lt~FP16 & 523 & 92.5 & 139 & 98.0 & 3.77$\times$ \\
     & Lt~FP8 & 1,039 & 99.0 & \multicolumn{2}{c}{--} & 7.49$\times$ \\
     & rB~BF16 & 594 & 100 & 137 & 98.4 & 4.28$\times$ \\
     & rB~FP16 & 570 & 101 & 119 & 93.3 & 4.11$\times$ \\
    \bottomrule
  \end{tabular}
\end{table}

\subsubsection{Library capability gain}
hipBLASLt delivers most of FP8's speedup in large GEMMs and loses much of it in small ones. Fig.~\ref{fig:capture}(a) shows that the FP8 library capability gain ranges from \captureFPEightMin{} at $1024^3$ to \captureFPEightMax. The clock does not explain the loss: at the \captureFPEightNClocked{} shapes where we measured the FP8/FP16 clock ratio, dividing by it as well changes the capability gain by at most \captureDenomAgreementPct. The capability gain exceeds 1.0 at \captureFPEightNOverOneFixedLayout{} shapes ($2048^3$) because the FP16 kernel is weak. hipBLASLt's own heuristic does not always pick the best kernel: its top choice costs up to \selectionEffectSettled{} against the fastest kernel the library offers once the process has used the library, and in a new process it can be \selectionEffectWorstFactor{} slower than the best. The heuristic costs the MI250X less than its GCD spread.

\subsubsection{Cause of the loss}
hipBLASLt's kernel choice causes the loss at the MLP shape, the one shape we traced. Hardware counters show that hipBLASLt picks an FP8 tile twice as deep in $K$ as the FP16 one (\tileFPEightMlp{} against \tileFPSixteenMlp) and re-reads its operands \attribOperandRereadMlp{} as often. FP8 therefore moves \attribBytesPerFlopMlp{} FP16's bytes per FLOP, which cancels the benefit of its halved operand width. The same counters rule out conversion traffic, spilling, local data share (LDS), occupancy and clock.

\subsubsection{Generational speedup}
Upgrading to the MI300A and adopting FP8 together make an $8192^3$ GEMM \genCapPkgEndToEndGainEightK{} faster, package against package. We split this speedup into the FP16 generational speedup (CDNA3-FP16 over CDNA2-FP16 at the same configuration) and the FP8 library capability gain above. Fig.~\ref{fig:gencapture} compares their product with the measured end-to-end speedup. The FP16 generational speedup is \genCapPkgFactorOneGainRange.

\subsubsection{Per-CU\texorpdfstring{$\cdot$}{-}clock bound}
\begin{figure}[t]
  \centering
  \includegraphics[width=\columnwidth]{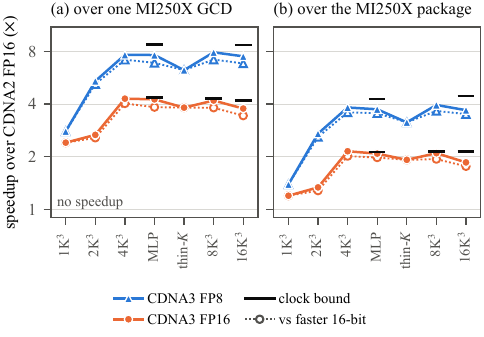}
  \caption{CDNA3 FP8 and FP16 GEMM speedup over CDNA2 FP16 on the MI250X package, per shape. Solid: against CDNA2 FP16; dotted, hollow: against its faster 16-bit path. Ticks: the per-CU$\cdot$clock bound, where both clocks were measured.}
  \label{fig:gencapture}
\end{figure}

The FP16 generational speedup is close to what CDNA3's hardware predicts: it reaches \genCapPkgFactorOneCapBRange{} of the per-CU$\cdot$clock bound, so CU count, clock and the doubled matrix rate explain almost all of it, and FP8 end to end reaches \genCapPkgEndToEndCapBRange{} of its bound. The bound has to use measured clocks. The measured clock ratio, \genCapClockRatioMeasuredRange, is much closer to one than the peak ratio \isaPeakClockRatio, and operand values move the clock: zero operands run up to \gemmInflationSpanMax{} faster on the MI300A, and the MI250X driver hides a drop in its delivered clock. Fig.~\ref{fig:gencapture} marks the shapes where both clocks were measured; neither the one-GCD scope nor CDNA2's faster BF16 baseline (dotted) changes these conclusions.

\section{Structured Sparsity and Its Break-Even}
\label{sec:sparsity}

For RQ2, 2:4 loses its speedup at two boundaries: the instruction clock costs part of it, and hipSPARSELt, which keeps only \captureSparseMin--\captureSparseMax{} of the $2\times$ on the MI300A, loses the rest in keeping the matrix units supplied with data. A cost model built from each stage predicts pruning within \sparsePruneHoldoutMaxErr{} but not the break-even point, and hipSPARSELt's kernel choice adds an uncertainty of up to \sparseSelectFloor{} to every sparse-versus-dense ratio. The SMFMAC FP8 instruction runs, but hipSPARSELt supports 2:4 only for FP16, BF16 and INT8~\cite{hipsparselt_docs}, as Table~\ref{tab:gen-support} shows.

\subsubsection{Instruction level}
2:4 sparsity loses part of its expected $2\times$ even at the instruction level, without any library. One SMFMAC instruction performs twice the multiply-accumulates of the dense MFMA instruction with the same tile and precision in the same number of issue cycles (\smfmacArchGainRange{} over \smfmacPairs{} pairs), so the $2\times$ holds per cycle. The sparse instructions, however, run the MI300A at \smfmacClockRatioRange{} of the dense clock, so the delivered speedup is only \smfmacDeliveredGainRange. We did not measure power in these runs, so we cannot attribute the lower clock; power-limited dynamic voltage and frequency scaling (DVFS) under the APU's single power cap is one candidate. hipSPARSELt then loses more.

\subsubsection{Library speedup}
hipSPARSELt's sparse matmul is at best slightly faster than dense, and usually slower. Fig.~\ref{fig:capture}(b) shows its capability gain (Eq.~\eqref{eq:gain}) over the \captureSparseNShapes{} shapes we use to fit the model below: it ranges from \captureSparseMin{} to \captureSparseMax, and the sparse matmul wins at only \sparseShapesFasterFrac{} of them. The best case, an LLM projection-layer shape held out from the fit, reaches a median speedup of \sparseSelectMatchedMedian{} across independent processes.

\subsubsection{Cause of the loss}
The sparse kernels lose their speedup in keeping the matrix units supplied with data, not in the matrix units themselves or in the clock. Counters show that SMFMAC delivers \spclkKappaMedian{} of its $2\times$ per cycle the matrix unit is busy (\spclkKappaRange{} over \spclkShapes{} shapes). The sparse kernel, however, keeps the matrix pipe busy only \spclkUpsilonMedian{} as often as the dense hipBLASLt kernel for the same GEMM (\spclkUpsilonRange) and issues \spclkLdsRatioMedian{} the LDS instructions per FLOP. It runs at \spclkRhoRange{}$\times$ the dense kernel's clock at the \spclkClockShapes{} shapes where we measured both, so the clock does not cause the loss. The instruction-level loss, by contrast, comes from the lower clock, so the library loses its speedup for a different reason than the instructions do.

\begin{table}[t]
  \centering
  \caption{hipSPARSELt prune and compress times on the MI300A at shapes from the model fit, and the model's error (measured/model $-1$). Sparse/dense: matmul time ratio, median over $N$.}
  \label{tab:gen-sparsity-stages}
  \footnotesize
  \setlength{\tabcolsep}{3pt}
  \begin{tabular}{lrrrrr}
    \toprule
    & \multicolumn{2}{c}{Prune} & \multicolumn{2}{c}{Compress} & Sparse/ \\
    \cmidrule(lr){2-3}\cmidrule(lr){4-5}
    $M\times K$ & $\mu$s & error & $\mu$s & error & dense \\
    \midrule
    $3072\times3072$ & 147 & +0.9\% & 117 & +40.9\% & 0.912 \\
    $6144\times6144$ & 584 & +0.0\% & 342 & +3.2\% & 0.886 \\
    $4096\times11008$ & 693 & -0.6\% & 396 & +0.2\% & 0.917 \\
    $4096\times11008$ (BF16) & 699 & +0.3\% & 392 & -0.9\% & 0.840 \\
    $8192\times1024$ & 130 & +0.4\% & 111 & +51.2\% & 0.938 \\
    $1024\times16384$ & 267 & +2.9\% & 177 & +20.5\% & 1.05 \\
    $16384\times16384$ & 4,213 & +1.4\% & 2,745 & +16.6\% & 1.00 \\
    \bottomrule
  \end{tabular}
\end{table}

\subsubsection{Break-even model}
A cost model built from each stage predicts the pruning cost, but not the break-even point. Sparsity pays off only when the matmul runs on enough columns $N$ to cover the cost of pruning and compressing the weights once. Pruning and compression both scale with the weight size: $T_{\text{prune}}=a\,MK$ and $T_{\text{comp}}=b\,MK$, with $c=a+b$. Planning costs a constant $\gamma$, and each matmul runs at its measured peak rate, $T=2MNK/R$, with $R_{\text{de}}$ for dense and $R_{\text{sp}}$ for sparse. The break-even width $N^{*}$ is where the one-shot sparse total, which pays every stage once, equals the dense matmul:
\begin{equation}
  N^{*}=\frac{c+\gamma/MK}{2\,(1/R_{\text{de}}-1/R_{\text{sp}})},
  \label{eq:nstar}
\end{equation}
which is independent of $M$ and $K$ once $\gamma/MK$ is small.%
 We fit the constants on these \captureSparseNShapes{} shapes and test them, unchanged, on \sparseHoldoutCells{} held-out shapes, which Table~\ref{tab:gen-sparsity-stages} lists; a cell is one shape at one value of $N$.

\begin{figure}[t]
  \centering
  \includegraphics[width=0.8\columnwidth]{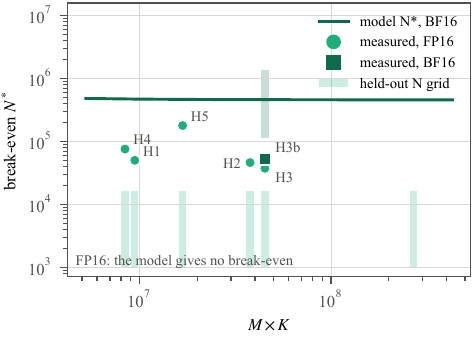}
  \caption{Measured 2:4 break-even crossings $N^{*}$ against the base model, by $MK$. Circles: FP16; squares: BF16; line: the model's BF16
  $N^{*}$; bars: held-out $N$ grids, with no crossing.}
  \label{fig:sparsity-breakeven}
\end{figure}

The model times pruning within \sparsePruneHoldoutMaxErr{} at every held-out shape but misses compression by up to \sparseCompressHoldoutMaxErr{} at the smallest $MK$, which points to a fixed compression cost it lacks. It also misses how the break-even depends on shape: Fig.~\ref{fig:sparsity-breakeven} shows that the measured FP16 crossings span \sparseExploratoryCrossingSpreadFPSixteen{} in $N^{*}$, which Eq.~\eqref{eq:nstar} cannot produce, and the model places the BF16 $N^{*}$ \sparseBFSixteenPredOverMeas{} above the measured crossing, because its constants assume a \sparseBFSixteenAdvPred{} per-FLOP advantage for sparse where the measured shape delivers \sparseBFSixteenAdvMeas. The FP16 ratio of the measured peaks, $R_{\text{sp}}/R_{\text{de}}=\sparseCeilQuotientFPSixteen$, predicts no break-even at all, yet the sparse matmul alone is faster at \sparseHoldoutFasterFracFPSixteen{} held-out cells, because the two libraries reach their measured peaks at different shapes. Users therefore have to measure a sparse-versus-dense ratio per shape.

hipSPARSELt also picks different kernels in different processes, at \sparseSelectNondetCells{} cells even with the node and binary held equal. We call the largest spread of the end-to-end ratio across such processes the \emph{selection floor}, \sparseSelectFloor; it makes every sparse-versus-dense ratio uncertain by that much, so in \sparseSelectFlipCount{} held-out FP16 cells where sparse measured faster, the margin is too small to call.%

\section{Capability Gain Under Sharing}
\label{sec:cotenancy}

For RQ3, the boundary is the hardware-queue pool: it, not precision, sets fairness on a shared device: with enough queues the pool treats every precision equally, and the default four queues trade fairness for throughput. FP8's speedup over FP16 grows when streams share queues, and the growth tracks a larger clock drop for FP16. 2:4 sparsity also tends to gain under sharing, and hard partitions isolate streams better than CU masks.

\subsubsection{Metrics}
Three metrics compare each stream's shared wall time with its solo time. Jain's index $J$~\cite{jain_fairness_1984} measures how evenly sharing slows the streams; it is 1 when all slow down equally, and we compare it only across streams that run the same GEMM. System throughput (STP) sums each stream's solo-over-shared time, so it reaches $n$ for $n$ perfectly overlapped streams and is 1 when sharing is no faster than running them one after another. Average normalized turnaround time (ANTT) is the mean shared-over-solo time, that is, the average slowdown.

\begin{table}[t]
  \centering
  \caption{Fairness and throughput of concurrent streams by precision and stream count $n$ at $4096^3$; $J$, STP and ANTT with 16 hardware queues, $J_4$ with the default 4. $J$: Jain's fairness index; STP: system throughput; ANTT: average normalized turnaround time.}
  \label{tab:gen-cotenancy}
  \footnotesize
  \setlength{\tabcolsep}{2.2pt}
  \begin{tabular}{@{}lrrrrrrrrr@{}}
    \toprule
    & & \multicolumn{4}{c}{MI300A} & \multicolumn{4}{c}{MI250X GCD} \\
    \cmidrule(lr){3-6}\cmidrule(l){7-10}
    Streams & $n$ & $J$ & $J_4$ & STP & ANTT & $J$ & $J_4$ & STP & ANTT \\
    \midrule
    dense FP8 & 1 & 1.00 & 1.00 & 1.00 & 1.00 & -- & -- & -- & -- \\
     & 2 & 0.910 & 0.913 & 1.50 & 1.49 & -- & -- & -- & -- \\
     & 4 & 0.996 & 0.960 & 0.98 & 4.12 & -- & -- & -- & -- \\
     & 8 & 1.00 & 0.875 & 1.14 & 7.02 & -- & -- & -- & -- \\
    \midrule
    dense FP16 & 1 & 1.00 & 1.00 & 1.00 & 1.00 & 1.00 & 1.00 & 1.00 & 1.00 \\
     & 2 & 0.907 & 1.00 & 1.51 & 1.48 & 0.993 & 0.997 & 1.08 & 1.86 \\
     & 4 & 0.997 & 0.922 & 0.98 & 4.12 & 1.00 & 0.939 & 1.01 & 3.97 \\
     & 8 & 1.00 & 0.875 & 0.96 & 8.33 & 0.999 & 0.924 & 0.88 & 9.11 \\
    \midrule
    2:4 sparse FP16 & 1 & 1.00 & 1.00 & 1.00 & 1.00 & -- & -- & -- & -- \\
     & 2 & 0.988 & 0.992 & 1.31 & 1.56 & -- & -- & -- & -- \\
     & 4 & 1.00 & 0.886 & 1.06 & 3.79 & -- & -- & -- & -- \\
     & 8 & 1.00 & 0.942 & 1.01 & 7.93 & -- & -- & -- & -- \\
    \midrule
    dense BF16 & 1 & \multicolumn{4}{l}{\emph{not measured}} & 1.00 & 1.00 & 1.00 & 1.00 \\
     & 2 & \multicolumn{4}{l}{\emph{not measured}} & 0.999 & 0.999 & 1.04 & 1.93 \\
     & 4 & \multicolumn{4}{l}{\emph{not measured}} & 1.00 & 0.920 & 1.03 & 3.90 \\
     & 8 & \multicolumn{4}{l}{\emph{not measured}} & 1.00 & 0.943 & 1.04 & 7.66 \\
    \bottomrule
  \end{tabular}
\end{table}

\subsubsection{Fairness}
The number of hardware queues, not the numeric precision, decides how fairly concurrent streams share the device, and changing the queue count trades fairness for throughput. Table~\ref{tab:gen-cotenancy} shows that with 16 queues, from four streams up, $J$ is at or near one for every precision on both GPUs, ANTT grows roughly as $n$, and STP stays near one. At $n=2$ the two MI300A streams take turns, one at its solo speed, so $J$ is low and STP far below two. With the default 4 queues, both sizes we ran with eight streams lose fairness and gain throughput: on the MI300A, $J$ falls to 0.875--0.942 at $4096^3$ and 0.927--0.938 at $8192^3$, while FP16 STP rises from 0.96 to 2.01 and from 0.63 to 1.12. The MI250X GCD shows the same trade.

\subsubsection{FP8}
FP8's speedup over FP16 grows when streams share hardware queues. The MI300A's FP16-over-FP8 wall-time ratio at $8192^3$ with 16 queues rises from \shareFPEightGainSolo{} alone to \shareFPEightGainEight{} at eight streams (\shareFPEightGainDelta), and the full ranges, \shareFPEightRangeSolo{} and \shareFPEightRangeEight, do not overlap. Capability gain (Eq.~\eqref{eq:gain}) rises from \shareFPEightCaptureSolo{} alone to \shareFPEightCaptureTwo{} at two streams and is still \shareFPEightCaptureEight{} at eight, so something other than arithmetic now drives the speedup. Queue sharing causes the growth: it takes the ratio from \isoQueueFPEightGainSolo{} alone to \isoQueueFPEightGainSix{} at six streams, while CPX partitions (one stream per XCD) leave it flat, \isoPartFPEightGainSolo{} to \isoPartFPEightGainSix.

The extra speedup tracks the clock. Power management under the APU's single power cap throttles FP16 harder than FP8, so the FP8/FP16 clock ratio grows from \clockRatioSolo{} alone to \clockRatioEight{} at eight streams, which explains \clockExplainedPct{} of the growth in capability gain; a second node repeats this. The attribution is consistent with DVFS but, with \clockSustainRepeats{} repeats per configuration on two nodes, does not prove it. The power cap does not explain everything: with eight streams at $2048^3$ the socket stays below its cap, yet FP16 still clocks lower than FP8.

\subsubsection{2:4 sparsity}
Sharing tends to favour 2:4 sparsity, but only the direction is reliable. Sparse runs about as fast as dense FP16 alone at $8192^3$ and \shareSparseGainEight{} faster at eight streams, and sharing favours sparse at \sparseShareTowardSparseCount{} shapes. Swapping only the sparse algorithm moves the eight-stream ratio by \sparseShareAlgoSwapEffect, and the selection floor of Section~\ref{sec:sparsity} is \sparseSelectFloor, so we do not report the size of the effect.

\subsubsection{Dividing the GPU between streams}
Hard partitions isolate streams best, and CU masks worst. Partitions deliver \isoPoverQnThree{} and \isoPoverQnSix{} the total throughput of queue sharing for three and six FP16 streams, while CU masks deliver only \isoMoverQnThree{} and \isoMoverQnSix. A mask cannot confine a stream to fewer than \xcdCount{} XCDs and fails silently, so masked streams still share every L2; we hypothesize that partitions win by giving each stream whole XCDs.%
 An administrator who shares an MI300A between jobs should therefore partition it rather than use CU masks.

\section{Application Case Studies}
\label{sec:apps}

To answer RQ4, how much of FP8's speedup reaches applications depends on fusion. FP8 speeds up the LLM only where its conversion is fused (vLLM serving, \appServeFPEightGainDynamicRange$\times$), not in our unfused harness, and a training step by only \appTrainFPEightGainMIThreeHundredA$\times$; 2:4 speeds up no layer; and on the MI250X the FP16 LU of rocHPL-MxP does no work, so that benchmark yields no generational ratio.

\begin{table}[t]
  \centering\footnotesize
  \caption{Llama-3.1-8B speedups on the MI300A. FP8: over FP16 (BF16 for training), static (s) or dynamic (d) scale. Gen.: FP16 (BF16) over one MI250X GCD and the package.}
  \label{tab:fp8-llm}
  \setlength{\tabcolsep}{3pt}
  \begin{tabular}{@{}llrrrr@{}}
    \toprule
    & & \multicolumn{2}{c}{FP8} & \multicolumn{2}{c}{Gen.} \\
    \cmidrule(lr){3-4}\cmidrule(l){5-6}
    Workload & Part & s & d & /GCD & /pkg. \\
    \midrule
    Harness & prefill & 1.01--1.35 & 0.79--1.10 & 1.97--2.94 & -- \\
            & decode  & 0.88--1.03 & 0.59--0.88 & 1.23--2.20 & -- \\
    vLLM    & fixed trace & 1.17--1.38 & 1.13--1.44 & 2.26--2.96 & 1.36--1.73 \\
            & mixed trace & 1.16--1.32 & 1.13--1.42 & 2.27--2.82 & 1.39--1.65 \\
    Training & step & -- & \appTrainFPEightGainMIThreeHundredA & \appTrainGenGcd & \appTrainGenPkg \\
    \bottomrule
  \end{tabular}
\end{table}

\subsection{FP8 in LLM inference, serving and training}
\label{sec:apps:llm}
\label{sec:apps:train}

FP8 speeds up the LLM only where its conversion to FP8 is fused. Table~\ref{tab:fp8-llm} lists FP8's speedup and the MI300A's generational speedup for \appSparseModel{} in each mode; our plain PyTorch harness is slower than a production engine, but both GPUs run the same code, so the ratios are fair.

\subsubsection{Unfused inference}
Prefill exercises the matrix engines, and the MI300A's FP16 prefill is \appPrefillGcdSlowdownRange{} faster than one GCD's. Decode is bound by the CPU launching kernels, not by the GPU: at batch 1 a step takes \appDecodeStepOverWeightBound{} the time needed to stream the weights once. Fused attention speeds an FP16 request at batch 32 by \appFusedReqGainMIThreeHundredABThirtyTwo$\times$ on the MI300A and \appFusedReqGainGcdBThirtyTwo$\times$ on a GCD, with the same first token.

FP8 runs on the MI300A but gains little. PyTorch's OCP type \texttt{\appLLMFPEightAcceptedButBroken} returns non-finite results on the MI300A, whose FP8 instructions use the FNUZ encoding (Section~\ref{sec:background}). With a static scale, calibrated in advance, FP8 is \appLLMFPEightGainStaticPrefill{} faster at prefill and \appLLMFPEightGainStaticDecode{} at decode, a loss, and a dynamic scale, computed per call, lowers decode to \appLLMFPEightGainDynamicDecode. The FP8 kernel alone is \appLLMFPEightKernelGain{} faster at the model's layer shapes, but the whole layer delivers \appLLMFPEightModelGain: the harness converts activations to FP8 in separate kernels, at a median \appLLMQuantTaxMedian{} per call, and at \appLLMQuantCostlierCells{} cells the conversion alone costs more than the FP16 matmul it replaces. This cost is an upper bound on what a fused engine pays.

\subsubsection{Fused serving (vLLM)}
A serving engine fuses activation quantization into the preceding kernel and serves many requests at once. We serve the model with vLLM at several concurrency levels (Section~\ref{sec:method:recipes}). Here $S_\ell$ in Eq.~\eqref{eq:gain} is the engine's FP8-over-FP16 throughput speedup.

Fusing the quantization makes FP8 pay off at every load we serve. FP8 increases the MI300A's output throughput by \appServeFPEightGainDynamicRange$\times$ with a dynamic scale and \appServeFPEightGainStaticRange$\times$ with a calibrated static scale, capability gains of \appServeFPEightCaptureDynamicRange{} and \appServeFPEightCaptureStaticRange, whereas the unfused harness above loses at decode. In FP16 the MI300A serves \appServeGenGcdRange$\times$ one MI250X GCD's throughput and \appServeGenPkgRange$\times$ the package's, which we run with tensor parallelism over both GCDs. At 64 requests FP8 cuts the median time to first token from \appServeTtftMIThreeHundredAFixedSixtyFour{} to \appServeTtftFPEightDynMIThreeHundredAFixedSixtyFour\,ms.

\subsubsection{Training}
Inference loses FP8 at small, launch-bound GEMMs; training runs large GEMMs, where hipBLASLt keeps nearly all of FP8's instruction-level speedup (\captureFPEightMax{} at $16384^3$, Section~\ref{sec:fp8}). Section~\ref{sec:method:recipes} describes the training setup.

\begin{figure}[t]
  \centering
  \includegraphics[width=\columnwidth]{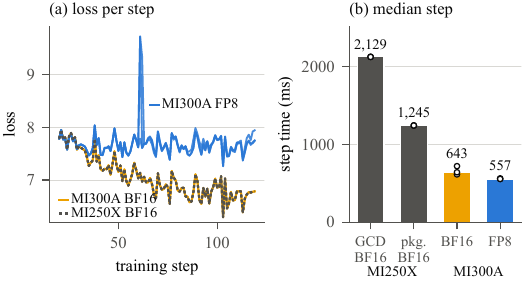}
  \caption{(a)~Loss per timed step, BF16 and FP8 on the MI300A and BF16 on one MI250X GCD (dotted). (b)~Median step time (bars) and each repeat's median (dots).}
  \label{fig:training}

      \vspace{-1em}
\end{figure}

Fig.~\ref{fig:training} shows that FP8 shortens a training step far less than its GEMM speedup predicts and that its loss stops falling. An FP8 step is \appTrainFPEightGainMIThreeHundredA$\times$ faster than BF16 and launches \appTrainKernelRatioFPEight$\times$ as many kernels, the same conversion cost as in inference. Its loss lies up to \appTrainLossFPEightMaxRelPctMIThreeHundredA{} above BF16's, so the speedup is per step, not per unit of training progress; we did not tune the FP8 recipe. In BF16 the MI300A step is \appTrainGenGcd$\times$ faster than one GCD's and \appTrainGenPkg$\times$ faster than the package's, and fused attention speeds it by a further \appTrainFusedGainMIThreeHundredA$\times$.

\subsubsection{FP8 accuracy}
FP8 costs little accuracy. We measure WikiText-2 perplexity, roughly the number of equally likely next tokens the model chooses among, so lower is better. Dense FP16 scores \accPplFPSixteenMIThreeHundredA{} on the MI300A and \accPplFPSixteen{} on one GCD, \accCrossFPSixteenPct\% apart, and fused attention changes the GCD's by \accPplFusedDeltaPctGcd\% and the MI300A's by \accPplFusedDeltaPctMIThreeHundredA\%. The per-tensor FP8 layers above increase perplexity by \accDeltaFPEightDynamicPctMIThreeHundredA\% \accDeltaFPEightDynamicPctMIThreeHundredACI{} with a dynamic scale and \accDeltaFPEightStaticPctMIThreeHundredA\% \accDeltaFPEightStaticPctMIThreeHundredACI{} with a static one. An FP8 inference or training stack therefore has to fuse the conversion before FP8's GEMM speedup can show.

\begin{table*}[t]
  \centering\footnotesize
  \caption{MI300A performance summary across workloads.}
  \label{tab:summary}
  \setlength{\tabcolsep}{3.5pt}
  \begin{tabular}{|l|l|r|r|r|l|c|}
    \hline
    \textbf{Workload} & \textbf{Metric} & \textbf{MI300A} & \textbf{FP8 or 2:4} & \textbf{vs.\ MI250X} & \textbf{Limiting factor} & \textbf{Sec.} \\
    \hline
    FP16 MFMA & TFLOP/s & 769--822 & -- & \isaDeviceRatioFPSixteenSixteen--\isaDeviceRatioFPSixteenThirtyTwo & $2\times$ per CU$\cdot$cycle only for 32$\times$32 & \ref{sec:isa} \\
    FP8 MFMA & TFLOP/s & 1,520--1,596 & \isaArchRatio$\times$ per cycle & -- & none & \ref{sec:isa} \\
    2:4 SMFMAC & vs.\ dense & -- & \smfmacDeliveredGainRange$\times$ & -- & lower clock & \ref{sec:sparsity} \\
    FP16 GEMM, hipBLASLt & measured peak & \ceilFPSixteenHipblaslt & -- & \genCapPkgFactorOneGainRange{} (pkg.) & CU count and clock & \ref{sec:fp8} \\
    FP8 GEMM, hipBLASLt & measured peak & \ceilFPEightHipblaslt & gain \captureFPEightMin--\captureFPEightMax & \genCapPkgEndToEndGainEightK{} (pkg.) & kernel choice & \ref{sec:fp8} \\
    2:4 GEMM, hipSPARSELt & capability gain & -- & \captureSparseMin--\captureSparseMax & -- & feeding the matrix units & \ref{sec:sparsity} \\
    Shared GEMM, 8 streams & FP8 over FP16 & -- & \shareFPEightGainEight{} (\shareFPEightGainSolo{} alone) & -- & queue pool and clock & \ref{sec:cotenancy} \\
    LLM prefill, batch 32 & time (ms) & 1,051 & layer \appLLMFPEightModelGain & \appPrefillGcdSlowdownBatchMax & unfused conversion & \ref{sec:apps:llm} \\
    vLLM, 64 requests & output tok/s & \appServeTokMIThreeHundredAFixedSixtyFour & \appServeFPEightGainDynamicRange$\times$ & 2.96$\times$, 1.73$\times$ & none (fused) & \ref{sec:apps:llm} \\
    LLM training step & time (ms), BF16 & \appTrainStepMsMIThreeHundredA & \appTrainFPEightGainMIThreeHundredA$\times$ & \appTrainGenGcd$\times$, \appTrainGenPkg$\times$ & conversion kernels & \ref{sec:apps:train} \\
    LLM layers, 2:4 & dense vs.\ sparse & -- & dense wins \appSparseOneShotDenseFaster & -- & library and pruning & \ref{sec:apps:sparse} \\
    Multi-process, $n{=}2$ & STP & \appTenantSTPTwoMIThreeHundredA & no FP8 speedup & within noise & little overlap & \ref{sec:apps:share} \\
    Partitioned serving, 6 proc. & CPX over SPX tok/s & \appPartRatioRange$\times$ & -- & per GCD \appPartRatioGcd$\times$ & launch-bound processes & \ref{sec:apps:share} \\
    rocHPL-MxP & time (s) & \hpcAppTMIThreeHundredAOneTwelveK & -- & not comparable & MI250X flushes FP16 & \ref{sec:apps:hpc} \\
    STREAM Triad & TB/s & \refWorkTriadMIThreeHundredA & -- & 3.13$\times$, 1.58$\times$ & HBM bandwidth & \ref{sec:apps:hpc} \\
    DGEMM, rocBLAS & TFLOP/s & \refWorkDgemmMIThreeHundredA & -- & 2.65$\times$, 1.37$\times$ & FP64 units & \ref{sec:apps:hpc} \\
    \hline
  \end{tabular}
\vspace{-2em}

\end{table*}

\subsection{2:4 sparsity in LLMs}
\label{sec:apps:sparse}

2:4 speeds up none of the model's layers and costs a lot of accuracy without fine-tuning.

\subsubsection{Sparse transformer layers}
2:4 does not speed up the model's linear layers on the MI300A, the only GPU with a 2:4 path, as Table~\ref{tab:gen-support} shows. The median dense/sparse ratio with weights compressed once and reused is \appSparseAmortMedian, inside the selection floor of Section~\ref{sec:sparsity} (\appSparseSpselFloorUsed): at that floor, \appSparseIndistinguishableSpselFloor{} cells are indistinguishable, \appSparseDenseFasterSpselFloor{} dense-faster and \appSparseSparseFasterSpselFloor{} sparse-faster. Dense wins at \appSparseOneShotDenseFaster{} cells when each call pays for compression.

\subsubsection{Accuracy of 2:4 pruning}
2:4 costs a lot of accuracy. Magnitude pruning of the \accDecoderLinears{} decoder linear layers increases perplexity from \accPplFPSixteenMIThreeHundredA{} to \accPplSMagDecMIThreeHundredA{} ($\times$\accRatioSMagDecMIThreeHundredA), and hipSPARSELt's own pruning pattern over all layers, the pattern timed above, to \accPplSTileAllMIThreeHundredA{} ($\times$\accRatioSTileAllMIThreeHundredA), which is close to random text. Usable 2:4 therefore needs an accuracy-aware pruner and fine-tuning, which we do not evaluate. The loss comes from pruning, not from hipSPARSELt's arithmetic, whose error matches torch's own FP16 product. Until hipSPARSELt keeps more of its $2\times$ and pruning keeps accuracy, dense remains the better choice for this model.

\subsection{LLM serving under sharing}
\label{sec:apps:share}

Whole models gain less from sharing a device than the GEMM streams of Section~\ref{sec:cotenancy}.

\subsubsection{Multi-process serving}
Several processes serving their own FP16 replicas overlap little: at $n=2$, $J$ is \appTenantJainTwo{} and STP is \appTenantSTPTwoMIThreeHundredA{} on the MI300A and \appTenantSTPTwoGcd{} on the GCD, out of 2, and the MI300A runs four processes only slightly better than serially (STP \appTenantSTPHomoFour{} with 4 queues, \appTenantSTPHomoFourHwqSixteen{} with 16). Unlike the GEMM streams, sharing does not turn FP8 into a speedup: an FP8 process alone takes \appTenantFPEightSoloQFour{} the FP16 process's time, and four FP8 processes take \appTenantFPEightMakespanQFour{} the time four FP16 processes take to finish with 4 queues and \appTenantFPEightMakespanQSixteen{} with 16.

\subsubsection{Partitioned serving}
At the GEMM level, hard partitions beat queue sharing and CU masks (Section~\ref{sec:cotenancy}). We compare CPX against SPX with whole models by aggregate throughput, per-process latency and $J$ (Section~\ref{sec:method:recipes}).

\begin{table}[t]
\centering\footnotesize
\caption{Partitioned against shared serving.}
\label{tab:gen-partitioned}
\setlength{\tabcolsep}{4pt}
\begin{tabular}{@{}lrrr@{}}
\toprule
Processes & partitioned & shared & ratio \\
\midrule
\multicolumn{4}{@{}l}{\emph{MI300A, 6 processes: one per CPX partition vs all on SPX}} \\
FP16 & 68.8 & 95.9\,/\,96.6 & 0.85\,/\,0.78 \\
FP8 & 88.3 & 95.9\,/\,95.9 & 0.92\,/\,0.93 \\
3 FP16 + 3 FP8 & 89.0 & 92.7\,/\,94.7 & 0.96\,/\,0.94 \\
\multicolumn{4}{@{}l}{\emph{MI250X, 2 processes: one per GCD vs both on one GCD}} \\
FP16 & 101.5 & 57.8\,/\,57.4 & 1.76\,/\,1.77 \\
\bottomrule
\end{tabular}
\end{table}

For whole models the result reverses (Table~\ref{tab:gen-partitioned}): \xcdCount{} CPX partitions deliver \appPartRatioRange$\times$ the throughput of the same processes sharing SPX, although a request on one partition, with a sixth of the CUs, takes only \appPartSoloSlowdownCpx$\times$ as long as on the whole device; one process per MI250X GCD delivers \appPartRatioGcd$\times$ that of both on one GCD.

The GEMM-level advantage of FP8 under sharing does not carry over to whole models served side by side.

\subsection{HPC workloads}
\label{sec:apps:hpc}
\label{sec:apps:hplmxp}

\begin{table}[t]
\centering\footnotesize
\caption{Median of rocHPL-MxP time to solution and rates on the MI300A and one MI250X GCD. $C$: rocBLAS FP16 measured peak; $R_T$: the application's GEMM rate; IR share: fraction of time in iterative refinement; $n_\mathrm{IR}$: refinement steps.}
\label{tab:gen-hpcapp}
\setlength{\tabcolsep}{3.5pt}
\begin{tabular}{@{}rrrrrrr@{}}
\toprule
$N$ & $t$ (s) & $R$ (TF/s) & $\Phi{=}R/C$ & $\Psi{=}R_T/C$ & IR share & $n_\mathrm{IR}$ \\
\midrule
\multicolumn{7}{@{}l}{\emph{MI300A (whole device)}} \\
30,720 & 0.192 & 100 & 0.178 & 0.672 & 0.260 & 3 \\
61,440 & 0.732 & 211 & 0.373 & 0.682 & 0.164 & 3 \\
112,640 & 3.01 & 316 & 0.559 & 0.742 & 0.096 & 3 \\
138,240 & 5.12 & 344 & 0.608 & 0.753 & 0.076 & 3 \\
\midrule
\multicolumn{7}{@{}l}{\emph{MI250X, one GCD (FP16 LU flushed)}} \\
30,720 & 0.307 & 63.0 & -- & -- & 0.261 & 5 \\
61,440 & 1.46 & 106 & -- & -- & 0.164 & 5 \\
112,640 & 7.08 & 135 & -- & -- & 0.081 & 4 \\
\midrule
\multicolumn{7}{@{}l}{\emph{MI250X package (both GCDs at once, two solves; flushed)}} \\
30,720 & 0.308 & 126 & -- & -- & 0.260 & 5 \\
61,440 & 1.46 & 212 & -- & -- & 0.164 & 5 \\
112,640 & 7.09 & 269 & -- & -- & 0.081 & 4 \\
\bottomrule

\vspace{-3em}
\end{tabular}
\end{table}

rocHPL-MxP's LU trailing update is the rocBLAS FP16 matrix-core GEMM of our measured peaks, so the question is how much of the FP16 measured peak reaches time to solution (Section~\ref{sec:method:recipes} describes the setup). rocHPL-MxP has no FP8 or BF16 path, and we port none. Table~\ref{tab:gen-hpcapp} shows that every timed solve passed. The MI300A's rate rises with $N$ and reaches $\Phi=\hpcAppPhiMIThreeHundredAOneThirtyEightK$ of the rocBLAS FP16 measured peak at the largest size. The MI300A solves \hpcAppTTSRatioOneTwelveK$\times$ faster than one GCD at the largest common size, and the package delivers \hpcAppSPkg$\times$ one GCD's throughput.

On the MI250X the FP16 LU does no work. The benchmark scales its generated matrix so that at least \ftzSubnormalFracMin{} of the FP16 operands fall below FP16's smallest normal, and the MI250X's FP16 matrix instructions flush subnormal inputs to zero~\cite{rocblas_guide}: a diagnostic build shows that the trailing updates receive only zeros, and the solve still passes because FP64 refinement, preconditioned by the FP32 LU of the diagonal blocks, needs \ftzNirStockRange{} steps instead of \ftzNirMI. On the MI300A the FP16 LU does its work, since multiplying the same operands in FP32 reproduces its residuals to within \ftzRTwoExactOverMIMaxDevPct. We therefore draw no generational ratio from this application, and the MI250X's $\Phi=\hpcAppPhiGcdOneTwelveK$ and $\Psi=\hpcAppPsiGcdOneTwelveK$, above its measured peaks, are the rate of a GEMM on zero operands. A generational comparison with an FP16 benchmark must first check that its operands are representable in FP16 on both GPUs.

\begin{table}[t]
\centering\footnotesize
\caption{Triad bandwidth and rocBLAS DGEMM rate on the MI300A and the MI250X. Triad: $a=b+s\,c$ (FP32), DGEMM: best shape.}
\label{tab:gen-refwork}
\setlength{\tabcolsep}{3.5pt}
\begin{tabular}{@{}lrrrr@{}}
\toprule
 & \multicolumn{2}{c}{Triad, TB/s} & \multicolumn{2}{c}{DGEMM, TFLOP/s} \\
\cmidrule(lr){2-3}\cmidrule(lr){4-5}
 & max (\% peak) & plateau & peak run (\% peak) & sweep \\
\midrule
MI300A & 3.26 (61\%) & 3.20 & 69.4 (57\%) & 69.5 \\
MI250X GCD & 1.04 (64\%) & 0.98 & 26.1 (55\%) & 26.3 \\
\quad vs MI300A & 0.319 & 0.307 & 0.377 & 0.379 \\
MI250X pkg. & 2.06 (63\%) & 1.99 & 50.7 (53\%) & -- \\
\quad vs MI300A & 0.633 & 0.621 & 0.731 & -- \\
\bottomrule
\end{tabular}
\end{table}

\subsubsection{Reference workloads}
\label{sec:apps:ref}
Table~\ref{tab:gen-refwork} lists a STREAM Triad, the same operation as BabelStream's Triad~\cite{deakin_gpustream_isc16,deakin_babelstream_ijcse18}, and rocBLAS DGEMM. One GCD reaches \refWorkTriadGcdVsMI{} of the MI300A's Triad bandwidth and \refWorkDgemmGcdVsMI{} of its DGEMM rate, far more than its FP16 share in Table~\ref{tab:gen-ceilings}, so the MI250X is relatively stronger at FP64 and bandwidth than at FP16.

\section{Discussion}
\label{sec:discussion}

Table~\ref{tab:summary} summarizes every workload on the MI300A: its result, what FP8 or 2:4 adds, its improvement over the MI250X, and the factor that limits it. Our measurements tell each audience what to change.

\textbf{Library developers.} hipBLASLt's heuristic costs up to \selectionEffectSettled{} against the fastest kernel the library offers (Section~\ref{sec:fp8}), so kernel selection should rest on measurement. FP8 pays in an LLM only once its conversion is fused into the preceding kernel (Section~\ref{sec:apps:llm}), and hipSPARSELt keeps at most \captureSparseMax{} of 2:4's $2\times$ because its kernels do not keep the matrix units supplied with data (Section~\ref{sec:sparsity}).

\textbf{System administrators.} Partitions isolate jobs that share an MI300A, but serve whole LLMs with less total throughput than SPX sharing (Section~\ref{sec:apps:share}), while CU masks cannot confine a stream to fewer than \xcdCount{} XCDs (Section~\ref{sec:cotenancy}). The hardware-queue pool, not precision, sets fairness, and the default four queues trade fairness for throughput, so the pool size should follow whichever of the two matters.

\textbf{Users.} FP8 is worth adopting for large, throughput-bound GEMMs, and dense should stay the default over 2:4. On the MI250X, which has neither feature, FP16 or BF16 through hipBLASLt is the choice, and moving that work to the whole MI300A gains \genCapPkgFactorOneGainRange{} before FP8 is used.

\section{Conclusion}
\label{sec:conclusion}

Each loss we measured has a specific cause at a specific boundary between layers. We traced capability gain for FP8 and 2:4 sparsity on the MI300A against the MI250X, from instructions to applications and a shared device. FP8 keeps most of its speedup through the library, making an $8192^3$ GEMM \genCapPkgEndToEndGainEightK{} faster than FP16 on the MI250X package. In an LLM, an unfused conversion to FP8 uses up that speedup (\appLLMFPEightModelGain{} per layer), while fused serving keeps a gain at every load. 2:4 sparsity keeps at most about half of its speedup in hipSPARSELt, so dense wins. Under sharing, the hardware-queue pool sets fairness and FP8's GEMM capability gain rises above 1, but shared LLM serving gains nothing from FP8, and CPX partitions serve it with less throughput than SPX. Mixed-precision HPL yields no generational ratio, because the MI250X flushes its FP16 operands.

\bibliographystyle{IEEEtran}
\bibliography{references}

\begin{thebibliography}{10}
\providecommand{\url}[1]{#1}
\csname url@samestyle\endcsname
\providecommand{\newblock}{\relax}
\providecommand{\bibinfo}[2]{#2}
\providecommand{\BIBentrySTDinterwordspacing}{\spaceskip=0pt\relax}
\providecommand{\BIBentryALTinterwordstretchfactor}{4}
\providecommand{\BIBentryALTinterwordspacing}{\spaceskip=\fontdimen2\font plus
\BIBentryALTinterwordstretchfactor\fontdimen3\font minus
  \fontdimen4\font\relax}
\providecommand{\BIBforeignlanguage}[2]{{%
\expandafter\ifx\csname l@#1\endcsname\relax
\typeout{** WARNING: IEEEtran.bst: No hyphenation pattern has been}%
\typeout{** loaded for the language `#1'. Using the pattern for}%
\typeout{** the default language instead.}%
\else
\language=\csname l@#1\endcsname
\fi
#2}}
\providecommand{\BIBdecl}{\relax}
\BIBdecl

\bibitem{micikevicius_fp8}
\BIBentryALTinterwordspacing
P.~Micikevicius, D.~Stosic, N.~Burgess, M.~Cornea, P.~Dubey, R.~Grisenthwaite,
  S.~Ha, A.~Heinecke, P.~Judd, J.~Kamalu, N.~Mellempudi, S.~Oberman,
  M.~Shoeybi, M.~Siu, and H.~Wu, ``{FP8} formats for deep learning,'' 2022.
  [Online]. Available: \url{https://arxiv.org/abs/2209.05433}
\BIBentrySTDinterwordspacing

\bibitem{ocp_fp8_spec}
\BIBentryALTinterwordspacing
{Open Compute Project}, ``{OCP} 8-bit floating point specification ({OFP8}),
  revision 1.0,'' Open Compute Project, Tech. Rep., Jun. 2023. [Online].
  Available: \url{https://github.com/opencomputeproject/FP8}
\BIBentrySTDinterwordspacing

\bibitem{gale_structured_sparsity}
\BIBentryALTinterwordspacing
A.~Mishra, J.~A. Latorre, J.~Pool, D.~Stosic, D.~Stosic, G.~Venkatesh, C.~Yu,
  and P.~Micikevicius, ``Accelerating sparse deep neural networks,'' 2021.
  [Online]. Available: \url{https://arxiv.org/abs/2104.08378}
\BIBentrySTDinterwordspacing

\bibitem{hpl_mxp}
\BIBentryALTinterwordspacing
J.~Dongarra, P.~Luszczek, and Y.~Tsai, ``{HPL-MxP} mixed-precision benchmark,''
  Benchmark website. [Online]. Available: \url{https://hpl-mxp.org/}
\BIBentrySTDinterwordspacing

\bibitem{haidar_mixed_precision_ir_sc18}
\BIBentryALTinterwordspacing
A.~Haidar, S.~Tomov, J.~Dongarra, and N.~J. Higham, ``Harnessing {GPU} tensor
  cores for fast {FP16} arithmetic to speed up mixed-precision iterative
  refinement solvers,'' in \emph{SC18: International Conference for High
  Performance Computing, Networking, Storage and Analysis}.\hskip 1em plus
  0.5em minus 0.4em\relax Dallas, TX, USA: IEEE, Nov. 2018, pp. 603--613.
  [Online]. Available: \url{https://doi.org/10.1109/SC.2018.00050}
\BIBentrySTDinterwordspacing

\bibitem{top500_nov2024}
{TOP500}, ``{TOP500} list, november 2024,''
  \url{https://top500.org/lists/top500/2024/11/}, Nov. 2024.

\bibitem{top500_jun2026}
------, ``{TOP500} list, june 2026,''
  \url{https://top500.org/lists/top500/2026/06/}, Jun. 2026.

\bibitem{amd_cdna2_whitepaper}
\BIBentryALTinterwordspacing
\emph{Introducing {AMD} {CDNA} 2 Architecture}, Advanced Micro Devices, 2021,
  white paper. [Online]. Available:
  \url{https://www.amd.com/content/dam/amd/en/documents/instinct-business-docs/white-papers/amd-cdna2-white-paper.pdf}
\BIBentrySTDinterwordspacing

\bibitem{mi200_datasheet}
\BIBentryALTinterwordspacing
\emph{{AMD} {Instinct} {MI200} Series Accelerator Data Sheet}, Advanced Micro
  Devices, 2021. [Online]. Available:
  \url{https://www.amd.com/content/dam/amd/en/documents/instinct-tech-docs/instinct-mi200-datasheet.pdf}
\BIBentrySTDinterwordspacing

\bibitem{amd_mi300a_whitepaper}
\BIBentryALTinterwordspacing
\emph{Introducing {AMD} {CDNA} 3 Architecture}, Advanced Micro Devices, 2025,
  white paper. [Online]. Available:
  \url{https://www.amd.com/content/dam/amd/en/documents/instinct-tech-docs/white-papers/amd-cdna-3-white-paper.pdf}
\BIBentrySTDinterwordspacing

\bibitem{hanindhito_cdna3_family_sc25}
\BIBentryALTinterwordspacing
B.~Hanindhito and B.~Patel, ``Characterizing performance, power, and energy of
  {AMD} {CDNA3} {GPU} family,'' in \emph{Proceedings of the International
  Conference for High Performance Computing, Networking, Storage and Analysis
  (SC '25)}.\hskip 1em plus 0.5em minus 0.4em\relax ACM, 2025, pp. 905--934.
  [Online]. Available: \url{https://doi.org/10.1145/3712285.3759768}
\BIBentrySTDinterwordspacing

\bibitem{mall_is_open_sc25}
\BIBentryALTinterwordspacing
A.~Tee, N.~Curtis, N.~Wolfe, and D.~Wong, ``The {MALL} is open: Exploring
  shared caches and latency in {AMD} {CDNA} 3 {GPUs},'' in \emph{Proceedings of
  the SC '25 Workshops of the International Conference for High Performance
  Computing, Networking, Storage and Analysis}.\hskip 1em plus 0.5em minus
  0.4em\relax ACM, 2025, pp. 1110--1116. [Online]. Available:
  \url{https://doi.org/10.1145/3731599.3767487}
\BIBentrySTDinterwordspacing

\bibitem{mi300a_datasheet}
\BIBentryALTinterwordspacing
\emph{{AMD} {Instinct} {MI300A} {APU} Data Sheet}, Advanced Micro Devices,
  2025. [Online]. Available:
  \url{https://www.amd.com/content/dam/amd/en/documents/instinct-tech-docs/data-sheets/amd-instinct-mi300a-data-sheet.pdf}
\BIBentrySTDinterwordspacing

\bibitem{cdna3_isa_guide}
\BIBentryALTinterwordspacing
\emph{``{AMD} {Instinct} {MI300}'' Instruction Set Architecture Reference
  Guide}, Advanced Micro Devices, Aug. 2025, cDNA3 ISA; revision of 5 August
  2025. [Online]. Available:
  \url{https://www.amd.com/content/dam/amd/en/documents/instinct-tech-docs/instruction-set-architectures/amd-instinct-mi300-cdna3-instruction-set-architecture.pdf}
\BIBentrySTDinterwordspacing

\bibitem{amr_partitioning}
\BIBentryALTinterwordspacing
A.~Abouelmagd, D.~Boehme, S.~Brink, J.~Burmark, M.~McKinsey, A.~Skjellum, and
  O.~Pearce, ``{GPU} partitioning, power, and performance of the {AMD}
  {MI300A},'' in \emph{Proceedings of the Supercomputing Asia and International
  Conference on High Performance Computing in Asia Pacific Region (SCA/HPCAsia
  '26)}.\hskip 1em plus 0.5em minus 0.4em\relax ACM, 2026, pp. 215--227.
  [Online]. Available: \url{https://doi.org/10.1145/3773656.3773680}
\BIBentrySTDinterwordspacing

\bibitem{hsa_runtime_spec}
\BIBentryALTinterwordspacing
\emph{{HSA} Runtime Programmer's Reference Manual}, HSA Foundation, 2021.
  [Online]. Available: \url{https://hsafoundation.com/standards/}
\BIBentrySTDinterwordspacing

\bibitem{schieffer_amd_matrix_cores_ispass24}
\BIBentryALTinterwordspacing
G.~Schieffer, D.~{Ara{\'u}jo De Medeiros}, J.~Faj, A.~Marathe, and I.~Peng,
  ``On the rise of {AMD} matrix cores: Performance, power efficiency, and
  programmability,'' in \emph{2024 IEEE International Symposium on Performance
  Analysis of Systems and Software (ISPASS)}.\hskip 1em plus 0.5em minus
  0.4em\relax IEEE, 2024, pp. 132--143. [Online]. Available:
  \url{https://doi.org/10.1109/ISPASS61541.2024.00022}
\BIBentrySTDinterwordspacing

\bibitem{blackwell_microbench}
\BIBentryALTinterwordspacing
A.~Jarmusch and S.~Chandrasekaran, ``Microbenchmarking {NVIDIA}'s {Blackwell}
  architecture: An in-depth architectural analysis,'' in \emph{2026 IEEE
  International Parallel and Distributed Processing Symposium (IPDPS)}.\hskip
  1em plus 0.5em minus 0.4em\relax IEEE, 2026, pp. 1026--1036. [Online].
  Available: \url{https://doi.org/10.1109/IPDPS65963.2026.00087}
\BIBentrySTDinterwordspacing

\bibitem{ambati_mi300x_analysis_arxiv}
\BIBentryALTinterwordspacing
C.~Ambati and T.~Diep, ``{AMD} {MI300X} {GPU} performance analysis,'' 2025.
  [Online]. Available: \url{https://arxiv.org/abs/2510.27583}
\BIBentrySTDinterwordspacing

\bibitem{hopper_benchmark}
\BIBentryALTinterwordspacing
W.~Luo, R.~Fan, Z.~Li, D.~Du, Q.~Wang, and X.~Chu, ``Benchmarking and
  dissecting the {Nvidia} {Hopper} {GPU} architecture,'' in \emph{2024 IEEE
  International Parallel and Distributed Processing Symposium (IPDPS)}.\hskip
  1em plus 0.5em minus 0.4em\relax IEEE, 2024, pp. 656--667. [Online].
  Available: \url{https://doi.org/10.1109/IPDPS57955.2024.00064}
\BIBentrySTDinterwordspacing

\bibitem{demystify_gpu_micro}
\BIBentryALTinterwordspacing
H.~Wong, M.-M. Papadopoulou, M.~Sadooghi-Alvandi, and A.~Moshovos,
  ``Demystifying {GPU} microarchitecture through microbenchmarking,'' in
  \emph{2010 IEEE International Symposium on Performance Analysis of Systems \&
  Software (ISPASS)}, 2010, pp. 235--246. [Online]. Available:
  \url{https://doi.org/10.1109/ISPASS.2010.5452013}
\BIBentrySTDinterwordspacing

\bibitem{gpu_microbenchmarking}
\BIBentryALTinterwordspacing
X.~Mei and X.~Chu, ``Dissecting {GPU} memory hierarchy through
  microbenchmarking,'' \emph{IEEE Transactions on Parallel and Distributed
  Systems}, vol.~28, no.~1, pp. 72--86, 2017. [Online]. Available:
  \url{https://doi.org/10.1109/TPDS.2016.2549523}
\BIBentrySTDinterwordspacing

\bibitem{wahlgren_mi300a_upm}
\BIBentryALTinterwordspacing
J.~Wahlgren, G.~Schieffer, R.~Shi, E.~A. Le{\'o}n, R.~Pearce, M.~Gokhale, and
  I.~Peng, ``Dissecting {CPU-GPU} unified physical memory on {AMD} {MI300A}
  {APUs},'' in \emph{2025 IEEE International Symposium on Workload
  Characterization (IISWC)}.\hskip 1em plus 0.5em minus 0.4em\relax IEEE, Oct.
  2025, pp. 368--380. [Online]. Available:
  \url{https://doi.org/10.1109/IISWC66894.2025.00038}
\BIBentrySTDinterwordspacing

\bibitem{kuncham_one_memory_many_paths_ipdps26}
\BIBentryALTinterwordspacing
G.~K.~R. Kuncham, S.~Zhang, and D.~K. Panda, ``One memory-many paths: Early
  experiences with allocation and data copy strategies on {MI300A},'' in
  \emph{2026 IEEE International Parallel and Distributed Processing Symposium
  (IPDPS)}.\hskip 1em plus 0.5em minus 0.4em\relax IEEE, 2026, pp. 1012--1025.
  [Online]. Available: \url{https://doi.org/10.1109/IPDPS65963.2026.00086}
\BIBentrySTDinterwordspacing

\bibitem{antepara_roofline_superchips_sc25}
\BIBentryALTinterwordspacing
O.~Antepara, L.~Oliker, and S.~Williams, ``Roofline analysis of tightly-coupled
  {CPU-GPU} superchips: A study on {MI300A} and {GH200},'' in \emph{Proceedings
  of the SC '25 Workshops of the International Conference for High Performance
  Computing, Networking, Storage and Analysis}.\hskip 1em plus 0.5em minus
  0.4em\relax ACM, 2025, pp. 1205--1216. [Online]. Available:
  \url{https://doi.org/10.1145/3731599.3767497}
\BIBentrySTDinterwordspacing

\bibitem{schieffer_mi300a_interapu}
\BIBentryALTinterwordspacing
G.~Schieffer, J.~Wahlgren, R.~Shi, E.~A. Le{\'o}n, R.~Pearce, M.~Gokhale, and
  I.~Peng, ``Inter-{APU} communication on {AMD} {MI300A} systems via {Infinity}
  {Fabric}: A deep dive,'' in \emph{Proceedings of the International Symposium
  on Memory Systems (MEMSYS '25)}.\hskip 1em plus 0.5em minus 0.4em\relax ACM,
  2025, pp. 238--250. [Online]. Available:
  \url{https://doi.org/10.1145/3767110.3767130}
\BIBentrySTDinterwordspacing

\bibitem{otterness_amd_ecrts20}
\BIBentryALTinterwordspacing
N.~Otterness and J.~H. Anderson, ``{AMD} {GPUs} as an alternative to {NVIDIA}
  for supporting real-time workloads,'' in \emph{32nd Euromicro Conference on
  Real-Time Systems (ECRTS 2020)}, ser. Leibniz International Proceedings in
  Informatics (LIPIcs), vol. 165.\hskip 1em plus 0.5em minus 0.4em\relax
  Schloss Dagstuhl--Leibniz-Zentrum f{\"u}r Informatik, 2020, pp. 10:1--10:23.
  [Online]. Available: \url{https://doi.org/10.4230/LIPIcs.ECRTS.2020.10}
\BIBentrySTDinterwordspacing

\bibitem{otterness_amd_scheduling}
\BIBentryALTinterwordspacing
------, ``Exploring {AMD} {GPU} scheduling details by experimenting with
  ``worst practices'','' in \emph{Proceedings of the 29th International
  Conference on Real-Time Networks and Systems (RTNS '21)}.\hskip 1em plus
  0.5em minus 0.4em\relax ACM, 2021, pp. 24--34. [Online]. Available:
  \url{https://doi.org/10.1145/3453417.3453432}
\BIBentrySTDinterwordspacing

\bibitem{quach_eclip_islped25}
\BIBentryALTinterwordspacing
R.~Quach, Y.~Wang, A.~Jahanshahi, D.~Wong, and H.~Kim, ``{ECLIP}:
  Energy-efficient and practical co-location of {ML} inference on spatially
  partitioned {GPUs},'' in \emph{2025 IEEE/ACM International Symposium on Low
  Power Electronics and Design (ISLPED)}.\hskip 1em plus 0.5em minus
  0.4em\relax IEEE, 2025, pp. 1--7. [Online]. Available:
  \url{https://doi.org/10.1109/ISLPED65674.2025.11261793}
\BIBentrySTDinterwordspacing

\bibitem{rapid_serve}
\BIBentryALTinterwordspacing
A.~Masood, P.~Gaur, and N.~Jayasena, ``{RAPID-Serve}: Resource-efficient and
  accelerated {P/D} intra-{GPU} disaggregation,'' 2026. [Online]. Available:
  \url{https://arxiv.org/abs/2601.11822}
\BIBentrySTDinterwordspacing

\bibitem{structured_sparsity}
\BIBentryALTinterwordspacing
W.~Wen, C.~Wu, Y.~Wang, Y.~Chen, and H.~Li, ``Learning structured sparsity in
  deep neural networks,'' in \emph{Advances in Neural Information Processing
  Systems 29 (NIPS 2016)}, 2016, pp. 2074--2082. [Online]. Available:
  \url{https://arxiv.org/abs/1608.03665}
\BIBentrySTDinterwordspacing

\bibitem{2_4_sparse_training}
\BIBentryALTinterwordspacing
Y.~Hu, K.~Zhao, W.~Huang, J.~Chen, and J.~Zhu, ``Accelerating transformer
  pre-training with 2:4 sparsity,'' in \emph{Proceedings of the 41st
  International Conference on Machine Learning (ICML)}, ser. Proceedings of
  Machine Learning Research, vol. 235, 2024, pp. 19\,531--19\,543. [Online].
  Available: \url{https://arxiv.org/abs/2404.01847}
\BIBentrySTDinterwordspacing

\bibitem{continuous_pruning_2_4}
\BIBentryALTinterwordspacing
Y.~Hu, J.~Zhu, and J.~Chen, ``{S-STE}: Continuous pruning function for
  efficient 2:4 sparse pre-training,'' in \emph{Advances in Neural Information
  Processing Systems 37 (NeurIPS 2024)}, 2024. [Online]. Available:
  \url{https://arxiv.org/abs/2409.09099}
\BIBentrySTDinterwordspacing

\bibitem{pytorch_2_4_sparsity}
\BIBentryALTinterwordspacing
J.~Cai, D.~Haziza, and S.~Rao, ``Accelerating neural network training with
  semi-structured (2:4) sparsity,'' PyTorch Blog, Jun. 2024. [Online].
  Available:
  \url{https://pytorch.org/blog/accelerating-neural-network-training/}
\BIBentrySTDinterwordspacing

\bibitem{dickmann_cug25}
\BIBentryALTinterwordspacing
D.~Dickmann, P.~Offenh{\"a}user, R.~Saxena, G.~Markomanolis, A.~Rigazzi,
  P.~Keller, K.~Kayabay, and D.~Hoppe, ``Evaluating {AMD} {Instinct} {MI300A}
  {APU}: Performance insights on {LLM} training via knowledge distillation,''
  in \emph{Proceedings of the Cray User Group (CUG '25)}.\hskip 1em plus 0.5em
  minus 0.4em\relax ACM, 2025, pp. 115--126. [Online]. Available:
  \url{https://doi.org/10.1145/3757348.3757361}
\BIBentrySTDinterwordspacing

\bibitem{vllm_sosp23}
W.~Kwon, Z.~Li, S.~Zhuang, Y.~Sheng, L.~Zheng, C.~H. Yu, J.~E. Gonzalez,
  H.~Zhang, and I.~Stoica, ``Efficient memory management for large language
  model serving with {PagedAttention},'' in \emph{Proceedings of the 29th
  Symposium on Operating Systems Principles (SOSP)}, 2023, pp. 611--626.

\bibitem{rochpl_mxp}
\BIBentryALTinterwordspacing
{Advanced Micro Devices}, ``{rocHPL-MxP}: {HPL-MxP} benchmark for {ROCm},''
  GitHub repository. [Online]. Available:
  \url{https://github.com/ROCm/rocHPL-MxP}
\BIBentrySTDinterwordspacing

\bibitem{hipsparselt_docs}
\BIBentryALTinterwordspacing
\emph{{hipSPARSELt} Documentation}, Advanced Micro Devices, 2026, accessed
  September 2026. [Online]. Available:
  \url{https://rocm.docs.amd.com/projects/hipSPARSELt/en/latest/}
\BIBentrySTDinterwordspacing

\bibitem{jain_fairness_1984}
\BIBentryALTinterwordspacing
R.~Jain, D.-M. Chiu, and W.~Hawe, ``A quantitative measure of fairness and
  discrimination for resource allocation in shared computer systems,'' Digital
  Equipment Corporation, Tech. Rep. DEC-TR-301, Sep. 1984. [Online]. Available:
  \url{https://arxiv.org/abs/cs/9809099}
\BIBentrySTDinterwordspacing

\bibitem{rocblas_guide}
\BIBentryALTinterwordspacing
\emph{{rocBLAS} Documentation}, Advanced Micro Devices, 2026, accessed
  September 2026. [Online]. Available:
  \url{https://rocm.docs.amd.com/projects/rocBLAS/en/latest/}
\BIBentrySTDinterwordspacing

\bibitem{deakin_gpustream_isc16}
\BIBentryALTinterwordspacing
T.~Deakin, J.~Price, M.~Martineau, and S.~McIntosh-Smith, ``{GPU-STREAM} v2.0:
  Benchmarking the achievable memory bandwidth of many-core processors across
  diverse parallel programming models,'' in \emph{High Performance Computing:
  ISC High Performance 2016 International Workshops, Revised Selected Papers},
  ser. Lecture Notes in Computer Science, M.~Taufer, B.~Mohr, and J.~M. Kunkel,
  Eds.\hskip 1em plus 0.5em minus 0.4em\relax Cham: Springer International
  Publishing, 2016, pp. 489--507. [Online]. Available:
  \url{https://doi.org/10.1007/978-3-319-46079-6_34}
\BIBentrySTDinterwordspacing

\bibitem{deakin_babelstream_ijcse18}
\BIBentryALTinterwordspacing
------, ``Evaluating attainable memory bandwidth of parallel programming models
  via {BabelStream},'' \emph{International Journal of Computational Science and
  Engineering}, vol.~17, no.~3, 2018. [Online]. Available:
  \url{https://doi.org/10.1504/IJCSE.2018.095847}
\BIBentrySTDinterwordspacing

\end{thebibliography}

\end{document}